\documentclass[pra,aps,twocolumn,superscriptaddress,floatfix,noshowpacs,nofootinbib,10pt]{revtex4-2}

\usepackage[utf8]{inputenc}
\usepackage{microtype}
\usepackage{amsmath,amssymb}
\usepackage{graphicx}
\usepackage[colorlinks=true]{hyperref}

\usepackage{bm}
\renewcommand{\vec}{\bm}

\renewcommand{\Im}{\operatorname{Im}}

\newcommand{\dif}{\mathrm{d}}
\newcommand{\mi}{\mathrm{i}}
\newcommand{\me}{\mathrm{e}}
\newcommand{\px}{s_p}
\newcommand{\unitvecx}{\hat{\vec{e}}_x}

\allowdisplaybreaks

\begin{document}

\title{Traveling supersolid stripe patterns in spin-orbit-coupled Bose-Einstein condensates}

\author{G. I. Martone}
\email{giovanni.martone@le.infn.it}
\affiliation{CNR NANOTEC, Institute of Nanotechnology, Via Monteroni, 73100 Lecce, Italy}
\affiliation{INFN, Sezione di Lecce, 73100 Lecce, Italy}

\author{G. V. Shlyapnikov}
\affiliation{Universit\'{e} Paris-Saclay, CNRS, LPTMS, 91405 Orsay, France}
\affiliation{Russian Quantum Center, Skolkovo, Moscow 143025, Russia}
\affiliation{Moscow Institute of Physics and Technology, Dolgoprudny, Moscow Region, 141701, Russia}
\affiliation{Van der Waals-Zeeman Institute, Institute of Physics, University of Amsterdam, Science Park 904, 1098 XH Amsterdam, The Netherlands}

\date{\today}

\begin{abstract}
We consider a traveling supersolid stripe pattern in a spin-orbit-coupled Bose gas. This configuration is associated with an unequal
population of the two single-particle energy minima, giving rise to a chemical potential difference that sets the fringe velocity.
Unlike stationary stripes, the moving pattern is spin-polarized, with decreasing contrast as the population imbalance increases,
eventually leading to stripe melting and transition to the uniform plane-wave phase. The Bogoliubov spectrum of the moving stripes
exhibits asymmetry under inversion of the excitation quasimomentum. At high population imbalance, we identify energetic and
dynamical instabilities in the spin-phonon mode which transforms to the roton mode of the plane-wave phase as the stripe structure
vanishes.
\end{abstract}

\maketitle

\section{Introduction}
\label{sec:intro}
In recent years, the interest in supersolidity, where superfluidity coexists with a crystal-like structure, has grown significantly
(see reviews in~\cite{Balibar_review,Boninsegni_review,Boettcher_review,Recati_review,Sinha_review,Chomaz_review}). First discussed
in the 1950s~\cite{Penrose1956,Gross1957,Gross1958} and further developed in subsequent decades~\cite{Thouless1969,Andreev1969,
Chester1970,Leggett1970,Kirzhnits1971,Saslow1975,Pitaevskii1984,Pomeau1994,Leggett1998}, the supersolidity has only recently been
observed in ultracold boson systems inside optical resonators~\cite{Leonard2017}, with spin-orbit coupling~\cite{Li2017,Putra2020,
Chisholm2024}, and with dipolar interactions~\cite{Tanzi2019,Boettcher2019,Chomaz2019,Norcia2021,Bland2021}, as well as in
exciton-polariton systems~\cite{Trypogeorgos2025,Muszynski2024}. Patterned states with supersolid-like dynamical features have also
been found in driven superfluids~\cite{Liebster2025}. These advances sparked studies on several aspects of supersolids,
including their ability to sustain a dissipationless flow, i.e., a supercurrent. At low flow velocities, the superfluid
fraction, bounded strictly below one due to crystalline order~\cite{Leggett1970,Leggett1998}, is inferred from the system's response
to a translation~\cite{Sepulveda2010,Roccuzzo2019,Martone2021a,Ancillotto2021,Ripley2023,Blakie2024,Orso2024} or rotation~\cite{
Josserand2007,Roccuzzo2020,Roccuzzo2022,Gallemi2022,Sindik2024,Preti2025} constraint. However, accessing this quantity in experiments
has proven challenging~\cite{Tanzi2021,Norcia2022}, and only very recently a measurement has been performed~\cite{Biagioni2024}. At
higher flow velocities, supercurrent-carrying supersolids can become unstable~\cite{Kunimi2012,Nilsson2021,Martone2021a,Mukherjee2025},
and in rotating configurations this can lead to the formation of vortices~\cite{Pomeau1994,Gallemi2020,Sindik2022,Schubert2025}, as
confirmed in recent experiments~\cite{Casotti2024,Poli2025}. An anomalous Doppler effect has also been predicted~\cite{Zawislak2025}.
Interestingly, although these results indicate that supersolids are (under certain conditions) able to sustain a supercurrent,
frictionless motion of external objects remains impossible~\cite{Pomeau1994,Martone2018}.

A common method to generate a supercurrent in a Bose-Einstein condensate (BEC) is to impose a phase twist on its time-independent order
parameter~\cite{Fisher1973}. In supersolids at zero temperature, this leads to a configuration where the normal component
associated with the lattice remains at rest, while the superfluid background flows at a velocity $\vec{v}_s$ determined by the
twist angle. An alternative approach involves considering a moving lattice pattern with velocity $\vec{v}_n = - \vec{v}_s$, making
the time-dependent order parameter a function of $\vec{r} - \vec{v}_n t$. In Galilean-invariant systems, such as dipolar gases,
these two descriptions are physically equivalent, being connected by a Galilean transformation. However, spin-orbit-coupled BECs lack
Galilean invariance, a feature that profoundly alters their superfluid behavior even in the non-supersolid phases. Early studies have
shown that in these phases the superfluid fraction is strictly less than unity even at zero temperature~\cite{Zhang2016,Chen2018,
Martone2021b}. In addition, the critical velocity for supercurrent stability differs from the threshold for frictionless impurity
motion~\cite{Zhu2012,Zheng2013}, and current-carrying configurations can exhibit both energetic and dynamical
instabilities~\cite{Ozawa2013}, unlike in standard condensates, where only energetic (Landau) instabilities are present~\cite{Landau1941,
Lifshitz_Pitaevskii_book}. In the supersolid stripe phase, the combined lack of translational and Galilean invariance further reduces the
superfluid density compared to the Leggett bound~\cite{Martone2021b} and leads to peculiar stability conditions for current-carrying
states~\cite{Lyu2024}.

\begin{figure}
\centering
\includegraphics[scale=1.0]{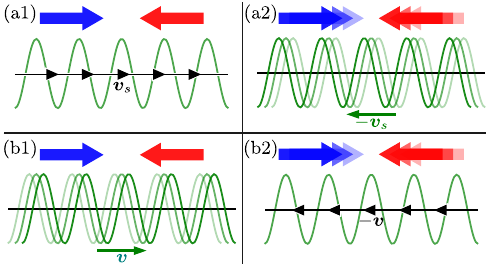}
\caption{Schematic illustration of background flow and lattice motion scenarios in the stripe phase of a spin-orbit-coupled BEC. (a1)
Supercurrent state in the laboratory frame: the superfluid background (horizontal black line) flows, while the Raman lasers generating the
spin-orbit coupling (blue and red arrows) and the density modulations (green wavy line) are at rest. (a2) The same supercurrent state in the
rest frame of the superfluid, where the Raman lasers and the fringes translate together. (b1) Traveling stripe pattern in the laboratory frame,
characterized by the absence of net mass transport at the coarse-grained level. (b2) Traveling stripe pattern in the comoving frame, where the
fringes are at rest (see Sec.~\ref{sec:exc_spectrum}). In all scenarios, the normal component is pinned to the Raman lasers.}
\label{fig:scenarios}
\end{figure}

The Raman lasers responsible for generating spin-orbit coupling play a crucial role in the construction of supercurrent states within the
stripe phase. In the analysis of Ref.~\cite{Lyu2024}, the current is generated by imposing twisted boundary conditions on the condensate
order parameter. This procedure effectively constrains the system's \textit{kinetic} momentum and describes a scenario in which the superfluid
background flows relative to the laboratory frame (i.e., the rest frame of the Raman lasers), while the density modulations remain stationary,
as shown in Fig.~\ref{fig:scenarios}(a1). This is consistent with the idea that the Raman lasers pin the normal component~\cite{Zhang2016}.
In the rest frame of the superfluid, both the fringes and the lasers appear to move at the same velocity [see Fig.~\ref{fig:scenarios}(a2)].
These configurations correspond to the two contrasting, but physically equivalent, pictures discussed earlier: background flow versus lattice
motion.

In this work, we unveil a third scenario, the one in which the density modulations move relative to the laboratory frame, with no net mass
transport at the coarse-grained (hydrodynamic) level, as illustrated in Fig.~\ref{fig:scenarios}(b1). This relative motion renders the
configuration physically distinct from the two previously discussed cases. Here, the relevant constrained quantity is the \textit{canonical}
momentum, which, unlike the kinetic momentum, commutes with the spin-orbit Hamiltonian~\cite{Ozawa2013}.

The traveling stripe patterns that fall within this third scenario can be understood as a generalization of the notion of an imbalanced
BEC mixture to the context of spin-orbit coupling, where the relevant degrees of freedom are dressed spin states rather than bare spin
components. Specifically, they emerge in the presence of a population imbalance between the two dressed spins entering the condensate
order parameter, that is, the two degenerate minima of the single-particle dispersion. This imbalance, quantified by a dimensionless variable
$-1 \leq \px \leq 1$ [see Eqs.~\eqref{eq:soc_stripe_spin_2h} and~\eqref{eq:soc_stripe_pop_imb} below for a more precise definition], leads
to the appearance of a chemical potential difference between the two states, which, in turn, drives the density fringes to move at a constant
velocity. Traveling stripes are therefore intermediate structures that smoothly interpolate between the stationary stripe phase, i.e., a
balanced mixture of two dressed spin states ($\px = 0$), and the uniform plane-wave phase corresponding to a fully polarized configuration
in dressed spin ($\px = \pm 1$). For infinitesimal $\px$ the fringe motion corresponds to the crystal Goldstone mode of the stripe phase
and can be excited by releasing a weak spin perturbation~\cite{Geier2021,Geier2023}. The goal of this work is to systematically study
traveling stripe configurations beyond the low-$\px$ regime and highlight their distinctive properties. We find that moving stripes have
higher energy and reduced contrast compared to stationary stripes. In addition, they exhibit asymmetric density profiles in the two bare
spin components and finite bare spin polarization. All these effects intensify with increasing population imbalance $\px$, till the fringes
disappear and the system turns into a uniform plane-wave condensate. Further differences with respect to stationary stripes emerge at the
level of dynamics. An analysis of the excitation spectrum of moving stripes reveals its asymmetry with respect to the center of the first
Brillouin zone. At sufficiently high $\px$ energetic and dynamical instabilities can emerge, which are not present in the stationary
ground-state stripe phase. These instabilities involve the spin phonon branch, that becomes the mode with the roton minimum of the
plane-wave phase in the $\px \to \pm 1$ limit~\cite{Martone2012}.

The mechanism considered here for generating a chemical potential difference through a population imbalance between dressed spin states relies
on the presence of antiferromagnetic spin-dependent interactions. The resulting fringe motion can thus be regarded as a purely interaction-driven
effect. A chemical potential mismatch can also be induced by an effective Zeeman splitting generated via finite Raman detuning, even when the two
dressed spin components in the order parameter have equal populations. Although this scenario is not explored in the present work, we note that
such a method was successfully used in the experiment of Ref.~\cite{Li2017} to observe moving stripes.

This paper is structured as follows. After briefly reviewing the model for a BEC with Raman-induced spin-orbit coupling
(Sec.~\ref{sec:model}), we outline the construction of traveling stripe solutions (Sec.~\ref{sec:trav_waves}). These solutions are then
derived and analyzed in detail (Sec.~\ref{sec:prop_mov_str}), with numerical results compared with perturbative analytical estimates.
Section~\ref{sec:exc_spectrum} focuses on the Bogoliubov spectrum and the instabilities of the moving stripes. The experimental feasibility
of observing these configurations is discussed in Sec.~\ref{sec:exp_persp}. Conclusions are presented in Sec.~\ref{sec:concl}.
Technical details of the perturbative approach and full expressions of relevant coefficients appearing in the formulas of the
main text are given in Appendices~\ref{sec:pert_method} and~\ref{sec:pert_formulas}, respectively.

\section{The model}
\label{sec:model}
Let us consider a spin-$1/2$ BEC with spin-orbit coupling. Within the mean-field approximation, the state of the system is described by a
two-component order parameter $\Psi(\vec{r},t) = (\Psi_\uparrow(\vec{r},t), \Psi_\downarrow(\vec{r},t))^T$, whose time evolution is
governed by the time-dependent Gross-Pitaevskii equation:
\begin{equation}
\mi \hbar \partial_t \Psi
= h_{\mathrm{SO}} \Psi + g_{dd} \left(\Psi^\dagger \Psi\right) \Psi + g_{ss} \left(\Psi^\dagger \sigma_z \Psi\right) \sigma_z \Psi \, .
\label{eq:soc_td_gp}
\end{equation}
Here, $h_{\mathrm{SO}}$ is the single-particle Hamiltonian incorporating spin-orbit coupling, and $\sigma_{x,y,z}$ the standard Pauli
matrices. The interaction parameters $g_{dd} = (g + g_{\uparrow\downarrow})/2$ and $g_{ss} = (g - g_{\uparrow\downarrow})/2$ represent
the density-density and spin-spin coupling strengths, respectively. In writing Eq.~\eqref{eq:soc_td_gp}, we have assumed equal
intraspecies interaction strengths, $g_{\uparrow\uparrow} = g_{\downarrow\downarrow} \equiv g$. The nonlinear coupling constants are
related to the corresponding $s$-wave scattering lengths through the standard relation $g_{\sigma\sigma'} = 4 \pi \hbar^2 a_{\sigma\sigma'}
/ m$ ($\sigma,\sigma' = \uparrow, \downarrow$), where $m$ is the atomic mass.

In the case of one-dimensional Raman-induced spin-orbit coupling, which is the focus of this work, the single-particle Hamiltonian
takes the form~\cite{Lin2011}
\begin{equation}
h_{\mathrm{SO}} = \frac{\left( p_x - \hbar k_R \sigma_z\right)^2}{2m} + \frac{p_y^2 + p_z^2}{2m}
+ \frac{\hbar\Omega_R}{2} \, \sigma_x + \frac{\hbar\delta_R}{2} \, \sigma_z \, ,
\label{eq:soc_ham}
\end{equation}
where $\vec{p} = - \mi \hbar \nabla_{\vec{r}}$ is the canonical momentum operator. The spin-orbit coupling strength is proportional to the
momentum imparted by the Raman lasers, which is $- 2 \hbar k_R \unitvecx$, with $\unitvecx$ the unit vector along the $x$ axis. The
corresponding energy scale is $E_R = \hbar^2 k_R^2 / 2m$. The parameter $\Omega_R$ quantifies the Raman coupling strength, while $\delta_R$
represents the Raman detuning, which we shall set to zero from now on.

It is worth noting that Eq.~\eqref{eq:soc_td_gp} can be recast in the variational form $\mi \hbar \partial_t \Psi = \delta E / \delta \Psi^\dagger$,
where $E$ is the total energy functional of the system:
\begin{equation}
E = \int_V \dif^3r \left( \Psi^\dagger h_{\mathrm{SO}} \Psi + \frac{g_{dd}}{2} n^2 + \frac{g_{ss}}{2} s_z^2 \right) \, .
\label{eq:soc_en}
\end{equation}
Here, $n = \Psi^\dagger \Psi$ is the total particle density and $s_z = \Psi^\dagger \sigma_z \Psi$ is the spin density along the $z$ axis. The
integral extends over the volume $V$ containing the condensate.

The two-component Gross-Pitaevskii equation~\eqref{eq:soc_td_gp} conserves two important quantities in addition to the total
energy~\eqref{eq:soc_en}. First, the invariance of the equation under global phase rotations of the order parameter $\Psi$ implies the
conservation of the particle number
\begin{equation}
N = \int_V \dif^3r \, \Psi^\dagger \Psi \, .
\label{eq:soc_part_numb}
\end{equation}
This condition corresponds to the standard normalization of $\Psi$.

Second, due to the space translation symmetry of Eq.~\eqref{eq:soc_td_gp}, the expectation value of the canonical momentum,
\begin{equation}
\langle \vec{p} \rangle = \int_V \dif^3r \, \Psi^\dagger \vec{p} \Psi \, ,
\label{eq:soc_can_mom}
\end{equation}
is conserved. On the other hand, the bare spin polarization, given by
\begin{equation}
\langle \sigma_z \rangle = \int_V \dif^3r \, \Psi^\dagger \sigma_z \Psi \, ,
\label{eq:soc_spin_pol}
\end{equation}
is conserved only if $\Omega_R = 0$, as the Raman coupling term in the single-particle Hamiltonian~\eqref{eq:soc_ham} does not
commute with $\sigma_z$. We point out that in the thermodynamic limit, where both the particle number and the system volume become
infinitely large, the conservation laws of $E$, $N$, $\langle \vec{p} \rangle$, and $\langle \sigma_z \rangle$ (for $\Omega_R = 0$)
are replaced by those of the corresponding average densities, which remain finite in this limit:
\begin{equation}
\bar{\varepsilon} = \frac{E}{V} \, ,
\quad \bar{n} = \frac{N}{V} \, ,
\quad \bar{\vec{p}} = \frac{\langle \vec{p} \rangle}{V} \, ,
\quad \bar{s}_z = \frac{\langle \sigma_z \rangle}{V} \, .
\label{eq:soc_cons_dens}
\end{equation}

We note that the canonical momentum $\vec{p}$ differs from the kinetic momentum $\vec{P} = \vec{p} - \hbar k_R \sigma_z \unitvecx$,
that is, the operator whose expectation value, $\langle \vec{P} \rangle = \int_V \dif^3r \, \Psi^\dagger \vec{P} \Psi$, gives the
center-of-mass velocity of the system multiplied by the total mass $N m$. One can also write $\langle \vec{P} \rangle = m \int_V \dif^3r
\, \vec{j}_d$, where
\begin{equation}
\vec{j}_d = \frac{\hbar}{2\mi m} \left( \Psi^\dagger \nabla_{\vec{r}} \Psi - \Psi^T \nabla_{\vec{r}} \Psi^* \right) - \frac{\hbar k_R}{m}
(\Psi^\dagger \sigma_z \Psi) \unitvecx
\label{eq:soc_part_curr}
\end{equation}
is the particle current density, that obeys the standard continuity equation~\cite{Martone2012}
\begin{equation}
\partial_t n + \nabla_{\vec{r}} \cdot \vec{j}_d = 0 \, .
\label{eq:soc_cont_eq}
\end{equation}
Due to the term $- \hbar k_R \sigma_z$ arising from the spin-orbit coupling, the $x$ component of the kinetic momentum does not commute
with the single-particle Hamiltonian~\eqref{eq:soc_ham}, unless the Raman coupling $\Omega_R$ vanishes. This implies the lack of Galilean
invariance in spin-orbit-coupled BECs, which has significant implications for the superfluid behavior, as discussed in the Introduction.
In the context of this work, the key implication is that constructing states with a constant superflow is not the same as constructing states
that propagate at a constant velocity. The former involves fixing the kinetic momentum $\langle\vec{P}\rangle$, as in the calculations of the
superfluid fraction in~\cite{Zhang2016,Chen2018} and in the study of current-carrying supersolid configurations in~\cite{Lyu2024}. In contrast,
here we focus on moving supersolid patterns with fixed canonical momentum $\langle \vec{p} \rangle$. For clarity we will henceforth refer to
this quantity simply as ``momentum,'' omitting the qualifier ``canonical.''

The ground state properties of an interacting spin-orbit-coupled BEC can be understood by first examining the structure of the single-particle
energy spectrum. Upon diagonalization of the Hamiltonian~\eqref{eq:soc_ham} at fixed momentum $\vec{p}$, two energy branches emerge. The lower
branch exhibits either two degenerate minima at momenta $\pm k_1^{\mathrm{SP}} \unitvecx = \pm k_R \sqrt{1 - (\hbar\Omega_R / 4 E_R)^2}
\unitvecx$ (for $\hbar\Omega_R < 4 E_R$) or a single minimum at zero momentum (when $\hbar\Omega_R \geq 4 E_R$)~\cite{Lin2011}.
The interplay between this structure and interaction effects gives rise to a rich equilibrium phase diagram. The phase structure can be inferred
by minimizing the total energy~\eqref{eq:soc_en} as a function of the system parameters~\cite{Ho2011,Li2012a} (for comprehensive overviews,
see also the reviews in Refs.~\cite{Zhou2013_review,Zhai2015_review,Li2015_review,Zhang2016_review,Martone2023_review}). Among the three quantum
phases that emerge, the stripe phase is of particular interest. This phase appears at low Raman coupling in the presence of an antiferromagnetic
spin-dependent interaction, i.e., when $g_{ss} > 0$. In this regime, both single-particle minima, corresponding to the dressed spin states
mentioned in the Introduction, are equally populated, resulting in a configuration with zero $\bar{s}_z$. This balanced occupation leads to the
formation of spatial modulations in the density profile. The emergence of such modulations indicates spontaneous breaking of translation symmetry.
When combined with the global $\mathrm{U}(1)$ phase symmetry breaking intrinsic to Bose-Einstein condensation, this reveals the supersolid nature
of the stripe phase. This interpretation is further corroborated by the properties of the Bogoliubov spectrum~\cite{Li2013,Martone2021b} (see also
Sec.~\ref{sec:exc_spectrum}).

In addition to the stripe phase, which is the main focus of this paper, the equilibrium phase diagram of spin-orbit-coupled BECs includes two
non-supersolid phases~\cite{Ho2011,Li2012a}. The plane-wave phase corresponds to a scenario where atoms occupy only one of the two
single-particle energy minima. Depending on which minimum is chosen, the momentum and bare spin polarization densities can take opposite values:
$\bar{\vec{p}} = \pm \bar{n} \hbar k_1^{\mathrm{PW}} \unitvecx$, $\bar{s}_z = \pm \bar{n} k_1^{\mathrm{PW}} / k_R$, where
\begin{equation}
k_1^{\mathrm{PW}} = k_R \sqrt{1 - \left(\frac{\Omega_R}{\Omega_{\mathrm{cr2}}}\right)^2}
\label{eq:soc_k1_pw}
\end{equation}
and $\Omega_{\mathrm{cr2}}$ is the critical Raman coupling defined below. In the plane-wave phase, the translation symmetry is unbroken,
and the density remains uniform and equal to the average density $\bar{n}$. This feature is shared by the second non-supersolid phase,
the single-minimum phase in which both the momentum and spin polarization are zero.

The non-supersolid phases appear at higher Raman coupling values, where the energy cost of density modulations becomes prohibitively high.
At low average density $\bar{n}$, the stripe and plane-wave phases are separated by a first-order transition, which occurs at a critical Raman
coupling $\hbar\Omega_{\mathrm{cr1}} = 4 E_R \sqrt{2 g_{ss} / (g_{dd} + 2 g_{ss})}$ in the $\bar{n} \to 0$ limit~\cite{Ho2011,Li2012a}. Conversely,
the second-order transition from the plane-wave to the single-minimum phase takes place at the larger critical value $\hbar\Omega_{\mathrm{cr2}}
= 2 (2 E_R - g_{ss} \bar{n})$. As $\bar{n}$ increases, the plane-wave phase becomes less favorable and eventually disappears. As a result, the
system can have a first order transition directly from the stripe phase to the single-minimum phase~\cite{Li2012a,SanchezBaena2020}.

\section{Traveling wave patterns}
\label{sec:trav_waves}
We will now demonstrate how to construct solutions to the Gross-Pitaevskii equation~\eqref{eq:soc_td_gp} that describe stripe patterns moving
with constant velocity. We first illustrate the emergence of traveling stripes using an approximate two-harmonic Ansatz for the order parameter
that has a transparent physical interpretation (Sec.~\ref{subsec:trav_stripe_ansatz_2h}). We then introduce a Bloch-wave Ansatz that solves
Eq.~\eqref{eq:soc_td_gp} exactly, and explore several of its key properties (Sec.~\ref{subsec:trav_stripe_ansatz_bloch}). Finally, in
Sec.~\ref{subsec:eval_ord_param}, we describe the numerical and perturbative methods used to calculate the order parameter of traveling stripes.

\subsection{Emergence of traveling stripes: two-harmonic Ansatz}
\label{subsec:trav_stripe_ansatz_2h}
The physical origin of traveling stripes can be understood by considering the following two-harmonic Ansatz for the condensate order parameter,
that represents an immediate generalization of the one introduced in earlier work~\cite{Ho2011,Li2012a} (see also Refs.~\cite{Wang2010,Wu2011}
for analogous proposals in Rashba spin-orbit-coupled systems) to describe the stationary stripe phase:
\begin{equation}
\Psi(\vec{r},t)
= \tilde{\Psi}_{+1} \me^{\mi k_+ x} \me^{- \mi \mu_+ t / \hbar} + \tilde{\Psi}_{-1} \me^{\mi k_- x} \me^{- \mi \mu_- t / \hbar} \, .
\label{eq:soc_stripe_wf_2h}
\end{equation}
This Ansatz describes a configuration where the atoms condense in a mixture of two dressed spin states, one with momentum $\hbar k_+ > 0$ and
the other with momentum $\hbar k_- < 0$, both along the $x$ direction. We denote by $\tilde{\Psi}_{\pm 1}$ the two-component spinor part of
the wave functions of the two dressed spins. Different from the fully stationary Ansatz of Refs.~\cite{Ho2011,Li2012a}, here we introduce two
distinct chemical potentials $\mu_+$ and $\mu_-$ for the two dressed spin states, so that the respective populations densities
$\tilde{\Psi}_{+1}^\dagger \tilde{\Psi}_{+1}$ and $\tilde{\Psi}_{-1}^\dagger \tilde{\Psi}_{-1}$ are individually conserved. Equivalently, one
can impose the conservation of the average particle density,
\begin{equation}
\bar{n} = \tilde{\Psi}_{+1}^\dagger \tilde{\Psi}_{+1} + \tilde{\Psi}_{-1}^\dagger \tilde{\Psi}_{-1} \, ,
\label{eq:soc_stripe_dens_2h}
\end{equation}
and of the dressed spin polarization density,
\begin{equation}
\bar{s} = \tilde{\Psi}_{+1}^\dagger \tilde{\Psi}_{+1} - \tilde{\Psi}_{-1}^\dagger \tilde{\Psi}_{-1} \, .
\label{eq:soc_stripe_spin_2h}
\end{equation}
The corresponding Lagrange multipliers are the mean chemical potential $\mu_d = (\mu_+ + \mu_-) / 2$ and the chemical potential semi-difference
$\mu_s = (\mu_+ - \mu_-) / 2$. For later convenience we also define the dimensionless population imbalance between the two dressed
spins as the ratio
\begin{equation}
\px = \frac{\bar{s}}{\bar{n}} \, .
\label{eq:soc_stripe_pop_imb}
\end{equation}
Notice that $- 1 \leq \px \leq 1$ and that the extreme values $\px = \pm 1$ correspond to the macroscopic occupation of a single dressed spin
state, that is, to a plane-wave configuration. On the other hand, the value $\px = 0$ defines a balanced coherent mixture of two dressed spins,
which is precisely the stripe phase at equilibrium. As we will see, in our spin-orbit-coupled BEC $\px$ plays a very important role, similar
to that of the bare spin imbalance $\bar{s}_z / \bar{n}$ in the absence of spin-orbit coupling. The two quantities actually coincide in the
limit of zero Raman coupling, where the dressed spin states approach their bare spin counterparts (see Sec.~\ref{subsec:zero_order}).

The conservation of $\bar{s}$ follows from the conservation of the average momentum density $\bar{\vec{p}}$. A simple calculation gives
\begin{equation}
\bar{p}_x = \hbar (k_c \bar{n} + k_1 \bar{s}) \, ,
\label{eq:soc_stripe_mom_dens}
\end{equation}
while $\bar{p}_y = \bar{p}_z = 0$ because the order parameter~\eqref{eq:soc_stripe_wf_2h} is independent of $y$ and $z$. In
Eq.~\eqref{eq:soc_stripe_mom_dens} we have introduced the combinations $k_c = (k_+ + k_-) / 2$ and $k_1 = (k_+ - k_-) / 2$. They correspond
to the (signed) length of the mean condensation wave vector, $\vec{k}_c = k_c \unitvecx$, and to the half-length of the stripe wave vector,
$2 \vec{k}_1 = 2 k_1 \unitvecx$. The latter fixes the spatial periodicity, equal to $\pi / k_1$, of the density modulations.

The values of the components of $\tilde{\Psi}_{\pm 1}$, as well as those of $k_\pm$ and $\mu_\pm$, can be determined minimizing the system's
energy. If one imposes $\mu_+ = \mu_-$ in Eq.~\eqref{eq:soc_stripe_wf_2h} and only fixes the value of $\bar{n}$, like in Refs.~\cite{Ho2011,
Li2012a}, one finds that, for $\delta_R = 0$ and $g_{\uparrow\uparrow} = g_{\downarrow\downarrow}$, the ground state has either $\px = 0$
(in the stripe phase) or $\px = \pm 1$ (in the plane-wave phase). Conversely, if the two chemical potentials are kept distinct, then
$\px$ can take any value between $-1$ and $1$. Remarkably, from the form of the order parameter~\eqref{eq:soc_stripe_wf_2h}, one
sees that, if $\mu_+ \neq \mu_-$, the density modulations travel along $x$ at constant velocity proportional to the chemical potential
difference:
\begin{equation}
v = \frac{\mu_s}{\hbar k_1} \, .
\label{eq:soc_stripe_vel}
\end{equation}
We thus conclude that a population imbalance between the two dressed spin components of the condensate order
parameter~\eqref{eq:soc_stripe_wf_2h} induces a translation motion of the density fringes. The aim of this paper is to study in detail the
properties of these moving stripes.

It is natural to expect that in the $\bar{n} \to 0$ limit the two dressed spin states entering the Ansatz~\eqref{eq:soc_stripe_wf_2h} approach
the two single-particle energy minima. Since these two minima have opposite momenta and are degenerate in energy one has $k_+ = - k_- =
k_1^{\mathrm{SP}}$ and $\mu_+ = \mu_-$ irrespective of the value of the population imbalance $\px$. As we will see below, at finite density and
for $\px \neq 0$ interactions can induce a mismatch in both the magnitude of the condensation wave vectors and the chemical potentials, thus
being at the origin of the stripe motion.

\subsection{Traveling Bloch-wave Ansatz}
\label{subsec:trav_stripe_ansatz_bloch}
Although the two-harmonic Ansatz~\eqref{eq:soc_stripe_wf_2h} is very useful for obtaining a simple qualitative description of the stripe phase,
in both stationary and moving configurations, it must be emphasized that it does not represent an exact solution of the time-dependent
Gross-Pitaevskii equation~\eqref{eq:soc_td_gp}, unless the nonlinear couplings $g_{dd}$ and $g_{ss}$ are zero. As shown in previous work on
stationary stripes~\cite{Li2013,Martone2021b}, interactions cause the presence, in the condensate order parameter, of higher-order harmonic
terms. The same is expected to happen in moving stripes. In this paper we will thus construct and make use of an improved Ansatz. For this
purpose we look for solutions of Eq.~\eqref{eq:soc_td_gp} that travel with a constant velocity $v$ along $x$ in the laboratory frame,
and that are uniform along $y$ and $z$. Such solutions have the standard form
\begin{equation}
\Psi(\vec{r},t) = \me^{-\mi \mu t / \hbar} \Psi_0(x - v t) \, ,
\label{eq:soc_gp_trav}
\end{equation}
where $\mu$ is the chemical potential and $\Psi_0$ is a function of the comoving coordinate $x - v t$. Substituting the
expression~\eqref{eq:soc_gp_trav} into the Gross-Pitaevskii equation~\eqref{eq:soc_td_gp} and transforming to the new independent variables
$x' = x - v t$ and $t' = t$ (i.e., the coordinates in the comoving frame, see Sec.~\ref{subsec:bogo_theory} below), we obtain the following
time-independent equation for $\Psi_0$:
\begin{equation}
\begin{split}
&{} h_{\mathrm{SO}} \Psi_0 + g_{dd} \left(\Psi_0^\dagger \Psi_0\right) \Psi_0 + g_{ss} \left(\Psi_0^\dagger \sigma_z \Psi_0\right) \sigma_z
\Psi_0 \\
&{} = (\mu + v p_x) \Psi_0 \, ,
\end{split}
\label{eq:soc_ti_gp_p}
\end{equation}
where now the spatial derivatives are with respect to $x'$. For $v = 0$, Eq.~\eqref{eq:soc_ti_gp_p} reduces to the standard time-independent
Gross-Pitaevskii equation, whose solutions represent the stationary states of the condensate, including the ground state. The $v$-dependent
term on the right hand side was previously included in Ref.~\cite{Ozawa2013} to construct and study plane-wave states with non-optimal momentum.

Having mapped our problem into a stationary one, we can now follow the same strategy as Refs.~\cite{Li2013,Martone2021b} and look for a solution
of Eq.~\eqref{eq:soc_ti_gp_p} in the form of a Bloch wave, describing a configuration with periodic density modulations along $x$:
\begin{equation}
\Psi_0(x') = \me^{\mi k_c x'} \sum_{\text{$\bar{m}$ odd}} \tilde{\Psi}_{\bar{m}} \me^{\mi \bar{m} k_1 x'} \, .
\label{eq:soc_stripe_wf_bl}
\end{equation}
This order parameter is given by the product of a plane wave with wave vector $\vec{k}_c = k_c \unitvecx$ (which can be identified with
the mean condensation wave vector appearing within the two-harmonic approximation, see Sec.~\ref{subsec:trav_stripe_ansatz_2h}) and of a
$\pi / k_1$-antiperiodic (and therefore $2 \pi / k_1$-periodic) function of $x'$. This function is conveniently represented as an expansion
in the basis of plane waves with wave vector equal to an half-integer multiple of the stripe wave vector $2 \vec{k}_1 = 2 k_1 \unitvecx$
[note that the sum in Eq.~\eqref{eq:soc_stripe_wf_bl} runs over both positive and negative odd integers]. The corresponding two-component
expansion coefficients are denoted by $\tilde{\Psi}_{\bar{m}}$. The two-harmonic Ansatz~\eqref{eq:soc_stripe_wf_2h} can be recovered from
Eqs.~\eqref{eq:soc_gp_trav}--\eqref{eq:soc_stripe_wf_bl} by retaining only the $\bar{m} = \pm 1$ terms associated with the occupation of
the two minima of the single-particle spectrum. In contrast, the additional $|\bar{m}| > 1$ harmonics arise due to the nonlinearity of the
Gross-Pitaevskii equation~\eqref{eq:soc_ti_gp_p} that $\Psi_0$ satisfies. The mean chemical potential can be expressed in terms of $\mu$
and $v$ as $\mu_d = \mu + \hbar k_c v$, while for the chemical potential semi-difference one has $\mu_s = \hbar k_1 v$, consistent with
Eq.~\eqref{eq:soc_stripe_vel}.

The average momentum density computed from the order parameter~\eqref{eq:soc_stripe_wf_bl} is still given by
Eq.~\eqref{eq:soc_stripe_mom_dens}, but the average particle and dressed spin polarization densities now read
\begin{equation}
\bar{n} = \sum_{\text{$\bar{m}$ odd}} \tilde{\Psi}_{\bar{m}}^\dagger \tilde{\Psi}_{\bar{m}}
\label{eq:soc_stripe_dens_bl}
\end{equation}
and
\begin{equation}
\bar{s} = \sum_{\text{$\bar{m}$ odd}} \bar{m} \tilde{\Psi}_{\bar{m}}^\dagger \tilde{\Psi}_{\bar{m}} \, ,
\label{eq:soc_stripe_spin_bl}
\end{equation}
respectively. Note that $\bar{s}$ enters the contribution of the antiperiodic part of the order parameter~\eqref{eq:soc_stripe_wf_bl}
to the average momentum density~\eqref{eq:soc_stripe_mom_dens}. We will take this contribution in the first Brillouin zone, i.e., between
$- \bar{n} \hbar k_1$ and $\bar{n} \hbar k_1$, which is always possible because $k_c$ is only defined up to a multiple of $2 k_1$.
Consequently, the population imbalance~\eqref{eq:soc_stripe_pop_imb} ranges between $-1$ and $1$ even with the new definitions of
$\bar{n}$ and $\bar{s}$.

In the stationary stripe phase, the result $\bar{p}_x = 0$  can be understood through symmetry considerations. The time-independent
Gross-Pitaevskii equation~\eqref{eq:soc_ti_gp_p} with $v = 0$ is invariant under the action of the two operators $\sigma_x \mathcal{P}$
and $\sigma_z \mathcal{T}$, which combine spin rotation with parity ($\mathcal{P}$) and time reversal ($\mathcal{T}$),
respectively.\footnote{In a spin-$1/2$ system, the time-reversal operator is $\mathcal{T} = \mi \sigma_y \mathcal{K}$ where $\mathcal{K}$
denotes complex conjugation.} Thus, the order parameter resulting from the application of either of these two operators to
Eq.~\eqref{eq:soc_stripe_wf_bl} remains a solution of Eq.~\eqref{eq:soc_ti_gp_p} and can differ from the original only by a global phase
rotation and spatial translation. This can happen only if $k_c = 0$ and the expansion coefficients fulfill, up to a phase factor, the
equality $\tilde{\Psi}_{-\bar{m}}^* = \sigma_x \tilde{\Psi}_{\bar{m}}$~\cite{Li2013,Martone2021b}. Using these properties, one can easily
show that $\bar{s}$, and thus $\bar{p}_x$, vanish when $v = 0$. Conversely, when the stripes are in motion, the action of
$\sigma_x \mathcal{P}$ and $\sigma_z \mathcal{T}$ on Eq.~\eqref{eq:soc_stripe_wf_bl} generates a stripe pattern moving with velocity
$-v$ along $x$. The corresponding momentum density~\eqref{eq:soc_stripe_mom_dens} is therefore nonzero and changes sign when one of the two
above symmetry operations is performed. This is achieved by simultaneously changing the sign of $k_c$ and $\bar{s}$.

The time-independent Gross-Pitaevskii equation~\eqref{eq:soc_ti_gp_p} corresponds to the stationarity condition of the functional
\begin{equation}
\begin{split}
\bar{\varepsilon}_{\mathrm{GC}} &{} = \bar{\varepsilon} - \mu \bar{n} - v \bar{p}_x \\
&{} = \bar{\varepsilon} - \mu_d \bar{n} - \mu_s \bar{s} \, .
\end{split}
\label{eq:soc_stripe_gc_en}
\end{equation}
This functional can be regarded as the grand-canonical energy density of the system, with a constraint on the average particle density
and another on the momentum or the dressed spin polarization density. Since $(\bar{n},\mu_d)$ and $(\bar{s},\mu_s)$ are pairs of conjugate
thermodynamic variables, they fulfill the standard thermodynamic relations
\begin{subequations}
\label{eqs:soc_en_dens_der}
\begin{align}
\mu_d = {}&{} \frac{\partial \bar{\varepsilon}}{\partial \bar{n}} \, ,
\label{eq:soc_en_dens_der_d} \\
\mu_s = {}&{} \frac{\partial \bar{\varepsilon}}{\partial \bar{s}} \, .
\label{eq:soc_en_dens_der_s}
\end{align}
\end{subequations}
It is also useful to consider the derivatives of the average energy density with respect to $k_c$ and $k_1$. They are given by
\begin{subequations}
\label{eqs:soc_stripe_curr}
\begin{align}
\bar{j}_{d,x}
= {}&{} \frac{\partial\bar{\varepsilon}}{\partial (\hbar k_c)} = \frac{\hbar}{m} (k_c \bar{n} + k_1 \bar{s} - k_R \bar{s}_z) \, ,
\label{eq:soc_stripe_curr_d} \\
\bar{j}_{s,x}
= {}&{} \frac{\partial\bar{\varepsilon}}{\partial (\hbar k_1)} = \frac{\hbar}{m} (k_c \bar{s} + k_1 \bar{s}_2 - k_R \bar{s}_{1z}) \, ,
\label{eq:soc_stripe_curr_s}
\end{align}
\end{subequations}
where $\bar{n}$ and $\bar{s}$ are as in Eq.~\eqref{eq:soc_stripe_dens_bl} and~\eqref{eq:soc_stripe_spin_bl}, respectively, and
\begin{subequations}
\label{eqs:soc_stripe_curr_coeff}
\begin{align}
\bar{s}_z = {}&{} \sum_{\text{$\bar{m}$ odd}} \tilde{\Psi}_{\bar{m}}^\dagger \sigma_z \tilde{\Psi}_{\bar{m}} \, ,
\label{eq:soc_stripe_sz} \\
\bar{s}_2 = {}&{} \sum_{\text{$\bar{m}$ odd}} \bar{m}^2 \tilde{\Psi}_{\bar{m}}^\dagger \tilde{\Psi}_{\bar{m}} \, ,
\label{eq:soc_stripe_s1} \\
\bar{s}_{1z} = {}&{} \sum_{\text{$\bar{m}$ odd}} \bar{m} \tilde{\Psi}_{\bar{m}}^\dagger \sigma_z \tilde{\Psi}_{\bar{m}} \, .
\label{eq:soc_stripe_s1z}
\end{align}
\end{subequations}
Specifically, Eqs.~\eqref{eqs:soc_stripe_curr} are the derivatives of the kinetic energy density, which is the only term that depends explicitly
on $k_c$ and $k_1$. Equation~\eqref{eq:soc_stripe_curr_d} can be rewritten as $\bar{j}_{d,x} = \frac{1}{V} \int_V \dif^3r \, j_{d,x}$, where
$j_{d,x}$ is the $x$ component of the particle current density~\eqref{eq:soc_part_curr}. Therefore, $\bar{j}_{d,x}$ is nothing else than the
average particle current density along $x$. On the other hand, the physical interpretation of $\bar{j}_{s,x}$ is immediate in the limit of zero
Raman coupling (see Sec.~\ref{subsec:zero_order}), where it reduces to the $x$ component of the average bare spin current density. We thus argue
that, at finite $\Omega_R$, $\bar{j}_{s,x}$ plays the role of the average dressed spin current density in the $x$ direction. As discussed in
Sec.~\ref{subsec:eval_ord_param} below, in this work we will only consider configurations with optimal values of $k_c$ and $k_1$, and thus with
zero $\bar{j}_{d,x}$ and $\bar{j}_{s,x}$. In particular, the vanishing of $\bar{j}_{d,x}$ entails that the momentum along $x$ is always
proportional to the bare spin polarization, that is,
\begin{equation}
\bar{p}_x = \hbar k_R \bar{s}_z \, .
\label{eq:soc_stripe_mom_spin}
\end{equation}
In addition, having $\bar{j}_{d,x} = 0$ and $\bar{j}_{s,x} = 0$ means that the stripe motion produces only oscillatory microscopic currents and
no macroscopic mass transport. We mention that, for stationary stripes, configurations with non-optimal $k_c$~\cite{Lyu2024} or $k_1$~\cite{Xia2023}
have been explored, and found to be stable when the two wave vectors are sufficiently close to the optimal values.

\subsection{Evaluation of the traveling stripe order parameter}
\label{subsec:eval_ord_param}
Having presented the structure and the general properties of the order parameter of traveling stripe configurations, we now discuss how it can be
explicitly determined. In the present work, we employ two complementary approaches. The first is a variational method, which begins inserting the
Ansatz~\eqref{eq:soc_stripe_wf_bl} into the total energy~\eqref{eq:soc_en} and performing the spatial integration. One obtains an expression for
the energy density $\bar{\varepsilon} = E / V$ as a function of $k_c$, $k_1$, and the two-component expansion coefficients $\tilde{\Psi}_{\bar{m}}$.
Using this result in combination with Eqs.~\eqref{eq:soc_stripe_dens_bl} and~\eqref{eq:soc_stripe_spin_bl} one immediately determines the
grand-canonical energy~\eqref{eq:soc_stripe_gc_en}. We then minimize this energy with the Lagrange multipliers $\mu_d$ and $\mu_s$ chosen so that
$\bar{n}$ and $\bar{s}$ take the desired values. As mentioned earlier, the stationarity condition of $\bar{\varepsilon}_{\mathrm{GC}}$ with respect
to $k_c$ and $k_1$ implies the vanishing of the average particle and dressed spin current densities, see Eqs.~\eqref{eqs:soc_stripe_curr}. To perform
the numerical minimization over the components of the $\tilde{\Psi}_{\bar{m}}$'s, the expansion~\eqref{eq:soc_stripe_wf_bl} must be truncated to a finite
number of harmonics. In our calculations, we retain terms with $- 9 \leq \bar{m} \leq 9$, which is sufficient to achieve high accuracy in all physical
quantities of interest. The validity of this variational method is confirmed by a direct comparison with the numerical solution of the Gross-Pitaevskii
equation~\eqref{eq:soc_ti_gp_p} in a periodic box. We find that the obtained wave function approaches the outcome of the variational calculation in the
limit of large simulation boxes.\footnote{We recall that, for a periodic box of length $L_x$ along $x$, the allowed values of $k_c$ and $k_1$ must
be integer multiples of $\pi / L_x$. The spacing between these values decreases as $L_x$ increases and vanishes in the limit $L_x \to \infty$. In this
limit, $k_c$ and $k_1$ become continuous variables, and their optimal values approach those obtained from the variational analysis.} In particular,
an examination of the momentum distribution reveals the presence of peaks associated with higher-order harmonics, thereby confirming the validity of
including these terms in the Ansatz~\eqref{eq:soc_stripe_wf_bl}.

In Ref.~\cite{Martone2021b}, a perturbative approach was developed to analytically compute the equilibrium order parameter in the stripe phase, along with
the associated observables. This method has proven highly accurate deeply in the double-minimum regime of the single-particle spectrum, which corresponds
to low Raman coupling, i.e., $\hbar\Omega_R / 4 E_R \ll 1$. In the present work, we extend this perturbative approach to investigate the case of a moving
striped condensate, thereby gaining deeper insight into the underlying physics. This method serves as an analytical alternative to the numerical
minimization discussed earlier. Notably, the two approaches yield excellent agreement for small enough values of the Raman coupling (for the values of the
interaction parameters we use in this work, we find that this agreement holds up to approximately $\hbar\Omega_R / E_R = 1.5$). We begin by considering
the power series expansion of the order parameter,
\begin{equation}
\Psi_0 = \Psi_0^{(0)} + \sum_{l = 1}^{+ \infty} \Psi_0^{(l)} \, ,
\label{eq:pert_exp_Psi}
\end{equation}
along with the expansions for the mean chemical potential,
\begin{equation}
\mu_d = \mu_d^{(0)} + \sum_{l = 1}^{+\infty} \mu_d^{(l)} \, ,
\label{eq:pert_exp_mu_n}
\end{equation}
and the chemical potential semi-difference,
\begin{equation}
\mu_s = \mu_s^{(0)} + \sum_{l = 1}^{+\infty} \mu_s^{(l)} \, .
\label{eq:pert_exp_mu_s}
\end{equation}
Here and in the following, the superscript ``$(l)$'' denotes the contribution of order $l$ in the small expansion parameter $\hbar\Omega_R / 4 E_R$. The
zero order terms in each series coincide with the results obtained in the absence of Raman coupling (see Sec.~\ref{subsec:zero_order} and
Appendix~\ref{sec:pert_method}). Substituting these expansions into the stationary Gross-Pitaevskii equation~\eqref{eq:soc_ti_gp_p} yields a set of
recurrence relations, which allow for the computation of each order-$l$ correction in terms of lower-order results up to $l - 1$. A detailed derivation
of this procedure, together with the expressions for the order parameter up to the second order in $\hbar\Omega_R / 4 E_R$, is given in
Appendix~\ref{sec:pert_method} and Appendix~\ref{sec:pert_formulas}.

\section{Properties of moving striped patterns}
\label{sec:prop_mov_str}
After establishing the general framework, we now proceed with a detailed analysis of the moving stripe patterns. We begin by examining the exact solution for the
case of zero Raman coupling, which can be mapped to an unbalanced BEC mixture (Sec.~\ref{subsec:zero_order}). Then, we present the results for several key
observables at finite Raman coupling (Sec.~\ref{subsec:res_obs}).

\subsection{Solution at zero Raman coupling. Unbalanced mixture}
\label{subsec:zero_order}
At zero Raman coupling ($\Omega_R = 0$), the lowest-energy solution of Eq.~\eqref{eq:soc_ti_gp_p} that satisfies the constraints~\eqref{eq:soc_stripe_dens_bl}
and~\eqref{eq:soc_stripe_spin_bl} has a simple analytical form:
\begin{equation}
\Psi_0^{(0)}(x') =
\begin{pmatrix}
\tilde{\Psi}_{+1,\uparrow}^{(0)} \\ 0
\end{pmatrix}
\me^{\mi k_R x'}
+
\begin{pmatrix}
0 \\ \tilde{\Psi}_{-1,\downarrow}^{(0)}
\end{pmatrix}
\me^{- \mi k_R x'} \, ,
\label{eq:zero_Psi}
\end{equation}
where $\tilde{\Psi}_{+1,\uparrow}^{(0)} = \sqrt{\bar{n}(1+\px)/2} \, \me^{\mi (\theta + \Delta\theta / 2)}$ and $\tilde{\Psi}_{-1,\downarrow}^{(0)} =
\sqrt{\bar{n}(1-\px)/2} \, \me^{\mi (\theta - \Delta\theta / 2)}$. This solution has $\mu_d^{(0)} = G_{dd}$ and $\mu_s^{(0)} = G_{ss} \px$, where
$G_{dd} = g_{dd} \bar{n}$ and $G_{ss} = g_{ss} \bar{n}$. Equation~\eqref{eq:zero_Psi} represents a special case of the stripe Ansatz~\eqref{eq:soc_stripe_wf_bl},
where only the $\bar{m} = \pm 1$ terms are nonzero. The optimal values of the wave vectors, $k_c^{(0)} = 0$ and $k_1^{(0)} = k_R$, result from energy
minimization. In contrast, the global phase of the order parameter, $\theta$, and the relative phase between the two spatially oscillating terms, $\Delta\theta$,
are not determined by energy minimization and can be chosen arbitrarily. The energy density for the configuration~\eqref{eq:zero_Psi} is $\bar{\varepsilon}^{(0)}
= \bar{n} (G_{dd} + G_{ss} \px^2) / 2$.

Equation~\eqref{eq:zero_Psi} describes a standard two-component BEC mixture with uniform particle density $\bar{n}$ and bare spin imbalance $\bar{s}_z^{(0)}
/ \bar{n} = \px$ (note that dressed spins reduce to their bare counterparts when $\Omega_R = 0$). This becomes even more evident when considering the full order
parameter, given by $\me^{- \mi \mu t / \hbar} \Psi_0^{(0)}(x - v t)$ [see Eq.~\eqref{eq:soc_gp_trav}], and performing the space-dependent spin rotation
$\exp(- \mi k_R \sigma_z x)$. This procedure transforms the order parameter into the canonical space-independent form\footnote{Applying the unitary transformation
$\exp(- \mi k_R \sigma_z x)$ to the single-particle Hamiltonian~\eqref{eq:soc_ham} with $\Omega_R = \delta_R = 0$ reduces it to the standard kinetic
energy $\vec{p}^2 / 2 m$. Consequently, Eq.~\eqref{eq:soc_td_gp} becomes the Gross-Pitaevskii equation for a two-component BEC without spin-orbit
coupling, and its uniform solutions take the form in Eq.~\eqref{eq:zero_Psi_mix}.}
\begin{equation}
\Psi_{\mathrm{mix}}(t) =
\begin{pmatrix}
\tilde{\Psi}_{+1,\uparrow}^{(0)} \me^{- \mi \mu_\uparrow t / \hbar} \\
\tilde{\Psi}_{-1,\downarrow}^{(0)} \me^{- \mi \mu_\downarrow t / \hbar}
\end{pmatrix} \, .
\label{eq:zero_Psi_mix}
\end{equation}
Notice that the combinations $\mu_{\uparrow,\downarrow} = \mu_d^{(0)} \pm \mu_s^{(0)} = G_{dd} \pm G_{ss} \px$ correspond exactly to the chemical potentials
of the two components in a binary BEC mixture with equal masses and symmetric intraspecies interactions~\cite{Pitaevskii_Stringari_book,Pethick_Smith_book}.
The global phase of the order parameter~\eqref{eq:zero_Psi_mix}, $\theta - \mu_d^{(0)} t / \hbar$, evolves in time at a rate determined by the mean chemical
potential, $\mu_d^{(0)} = (\mu_\uparrow + \mu_\downarrow) / 2$. The freedom in choosing the value of $\theta$ at $t = 0$ reflects the spontaneous breaking of
global $\mathrm{U}(1)$ phase symmetry, a hallmark of Bose-Einstein condensation. The polarization density vector of the mixture~\eqref{eq:zero_Psi_mix}
is defined as $\langle \vec{\sigma} \rangle_{\mathrm{mix}} / V = \left( \langle \sigma_x \rangle_{\mathrm{mix}}, \langle \sigma_y \rangle_{\mathrm{mix}},
\langle \sigma_z \rangle_{\mathrm{mix}} \right) / V$, where
\begin{subequations}
\label{eq:zero_pol_mix}
\begin{align}
\frac{\langle \sigma_x \rangle_{\mathrm{mix}}}{V} &{} = \bar{n} \sqrt{1 - \px^2} \cos \left[2 \mu_s^{(0)} t / \hbar - \Delta\theta\right] \, ,
\label{eq:zero_pol_mix_x} \\
\frac{\langle \sigma_y \rangle_{\mathrm{mix}}}{V} &{} = \bar{n} \sqrt{1 - \px^2} \sin \left[2 \mu_s^{(0)} t / \hbar - \Delta\theta\right] \, ,
\label{eq:zero_pol_mix_y} \\
\frac{\langle \sigma_z \rangle_{\mathrm{mix}}}{V} &{} = \bar{n} \px \, ,
\label{eq:zero_pol_mix_z}
\end{align}
\end{subequations}
and the subscript ``mix'' indicates that expectation values are evaluated with the order parameter~\eqref{eq:zero_Psi_mix}. When $\px \neq \pm 1$,
the polarization vector exhibits a nonzero component in the $xy$ plane, signaling the spontaneous breaking of spin rotational symmetry around the
$z$ axis. The azimuthal angle of the polarization is equal to minus the relative phase $\Delta\theta - 2 \mu_s^{(0)} t / \hbar$ between the two
components of the mixture~\eqref{eq:zero_Psi_mix}. At $t = 0$, the direction of the in-plane polarization is set by the arbitrary parameter
$-\Delta\theta$, and it subsequently rotates with a frequency determined by the chemical potential difference $2 \mu_s^{(0)} = \mu_\uparrow -
\mu_\downarrow$. This results in a precession of $\langle \vec{\sigma} \rangle_{\mathrm{mix}} / V$ around the $z$ axis, a phenomenon known in polariton
physics as self-induced Larmor precession~\cite{Shelykh2010}. This behavior originates from the $g_{ss}$-dependent term in the Gross-Pitaevskii
equation~\eqref{eq:soc_td_gp}, which effectively acts as a self-induced Zeeman splitting.

\begin{figure*}
\centering
\includegraphics[scale=1]{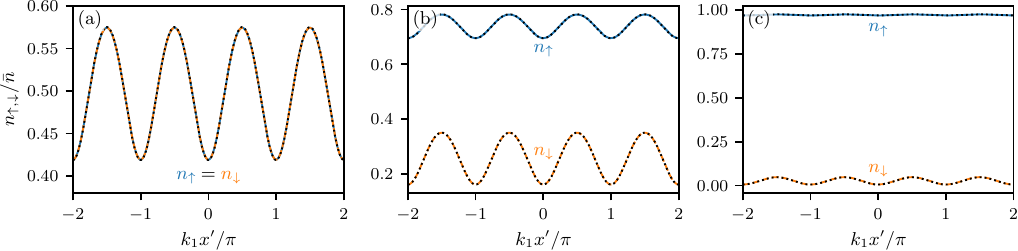}
\caption{Density profiles of spin-orbit-coupled BECs in the stripe phase for several values of the population imbalance between dressed spins:
(a) $\px = 0$ (stationary stripes), (b) $\px = 0.5$, and (c) $\px = 0.99$. The numerically obtained profiles of the bare spin-up (blue solid lines)
and spin-down (yellow dashed lines) components are shown. Close to each curve the prediction of the perturbative approach is also plotted (black dotted
lines). For definiteness, the fringe offset has been chosen such that a density minimum occurs at $x'=0$. The other parameters are $\hbar\Omega_R / E_R
= 1.0$, $G_{dd} / E_R = 1.2$, and $G_{ss} / E_R = 0.32$.}
\label{fig:dens_profiles}
\end{figure*}

\subsection{Results for the observables at finite Raman coupling}
\label{subsec:res_obs}
The discussion in the previous section demonstrates that, although the order parameter at zero Raman coupling $\Omega_R$, given by
Eq.~\eqref{eq:zero_Psi}, formally breaks translation invariance, this has no physical implications, since the spatial dependence can be eliminated
via a unitary transformation. However, the situation changes dramatically once the Raman coupling is introduced. On the one hand, due to the
$\sigma_x$-dependent term, the breaking of spin rotational symmetry around the $z$ axis becomes explicit. On the other hand, the spontaneous
breaking of translation symmetry now has observable consequences, most notably the emergence of density fringes. In this context, the
self-induced Larmor precession characteristic of the zero spin-orbit coupling regime is replaced at finite $\Omega_R$ by the self-induced
translational motion of the stripe pattern.

After evaluating the condensate order parameter at finite $\Omega_R$ using the two methods outlined in Sec.~\ref{subsec:eval_ord_param}, we now
discuss the behavior of several key observables. Throughout this and the following sections, we set the interaction parameters to $G_{dd} / E_R = 1.2$
and $G_{ss} / E_R = 0.32$ consistent with the values used in previous works~\cite{Li2013,Martone2021b}. With this choice of parameters, the first-order
transition separating the stripe and plane-wave phases in the ground-state phase diagram occurs at the critical Raman coupling $\hbar\Omega_{\mathrm{cr1}}
/ E_R = 2.70$. However, the stripe phase remains metastable up to the spinodal point $\hbar\Omega_{\mathrm{sp}} / E_R = 2.85$~\cite{Martone2021b}. In most
figures of this work, we compare results for two different values of the Raman coupling. The smaller value, $\hbar\Omega_R / E_R = 1.0$, lies within the
range of validity for the perturbative approach, where the predictions of this method match the numerical results excellently. The other value,
$\hbar\Omega_R / E_R = 2.6$, is so large that only the numerical method is applicable.

We first investigate the effect of stripe motion on the profiles of the total density and bare spin density. The total density is expressed as
\begin{equation}
n(x') = \bar{n} + \sum_{\bar{m} = 1}^{+ \infty} \tilde{n}_{\bar{m}} \cos[\bar{m} (2 k_1 x' + \Delta\theta)]
\label{eq:res_tot_dens}
\end{equation}
and the bare spin density is given by
\begin{equation}
s_z(x') = \bar{s}_z + \sum_{\bar{m} = 1}^{+ \infty} \tilde{s}_{z,\bar{m}} \cos[\bar{m} (2 k_1 x' + \Delta\theta)] \, .
\label{eq:res_spin_dens}
\end{equation}
As with the order parameter in Eq.~\eqref{eq:soc_stripe_wf_bl}, the total and spin densities are represented as Fourier series. These densities
oscillate around their respective average values, $\bar{n}$ and $\bar{s}_z$. The phase $\Delta\theta$ sets the offset of the density fringes.
Like in the case of $\Omega_R = 0$ (see Sec.~\ref{subsec:zero_order}), $\Delta\theta$ is randomly selected by the system following the spontaneous
breaking of translation invariance. The expansion coefficients $\tilde{n}_{\bar{m}}$ and $\tilde{s}_{z,\bar{m}}$ can be derived directly from those
of the order parameter (and the same applies for the bare spin polarization density $\bar{s}_z$, which is discussed below). At second order in the
Raman coupling, only the terms with $\bar{m} = 1$ and $\bar{m} = 2$ are nonzero. These coefficients are given by
\begin{subequations}
\label{eqs:res_coeffs}
\begin{align}
\frac{\tilde{n}_1}{\bar{n}} = {}&{} - \frac{8 E_R^3 (2 E_R+G_{ss}) \sqrt{1-\px^2}}{D_1} \frac{\hbar\Omega_R}{4 E_R} \, ,
\label{eq:res_dens_coeff_1} \\
\frac{\tilde{n}_2}{\bar{n}} = {}&{} - \frac{2 E_R^2 (1-\px^2) N_{\tilde{n}_2}}{D_1^2 D_2} \left( \frac{\hbar\Omega_R}{4 E_R} \right)^2 \, ,
\label{eq:res_dens_coeff_2} \\
\begin{split}
\frac{\tilde{s}_{z,1}}{\bar{n}} =
{}&{} 2 E_R \big[4 E_R^2 G_{ss} + 2 E_R \left(2 E_R+G_{ss}\right) G_{dd} \\
{}&{} - G_{ss}^3 \px^2\big] \frac{\px \sqrt{1-\px^2}}{D_1} \frac{\hbar\Omega_R}{4 E_R} \, ,
\end{split}
\label{eq:res_spin_coeff_1} \\
\frac{\tilde{s}_{z,2}}{\bar{n}} = {}&{} \frac{E_R^2 \px (1-\px^2) N_{\tilde{s}_{z,2}}}{D_1^2 D_2} \left( \frac{\hbar\Omega_R}{4 E_R} \right)^2 \, .
\label{eq:res_spin_coeff_2}
\end{align}
\end{subequations}
Here and in the other perturbative formulas presented in this section, the quantities $D_1$ and $D_2$ appearing in the denominators are even polynomial
functions of $\px$. Their full expressions are provided in Appendix~\ref{sec:pert_formulas}, see Eqs.~\eqref{eq:pert_form_D1} and~\eqref{eq:pert_form_D2}.
The same applies to the numerators $N_{\tilde{n}_2}$ and $N_{\tilde{s}_{z,2}}$, given in Eqs.~\eqref{eq:pert_form_N_n2} and~\eqref{eq:pert_form_N_s2},
respectively. It is worth noting that the total density $n$ is an even function of $\px$, while the bare spin density $s_z$ is an odd function.
In Fig.~\ref{fig:dens_profiles}, we show the combinations $n_{\uparrow,\downarrow} = (n \pm s_z) / 2$, corresponding to the densities of the two bare spin
components, for various values of $\px$. When $\px = 0$ [Fig.~\ref{fig:dens_profiles}(a)], the bare spin density vanishes, yielding equal profiles for the
two components: $n_\uparrow = n_\downarrow$. For $\px = 0.5$ [Fig.~\ref{fig:dens_profiles}(b)], the system becomes both globally and locally spin-polarized,
with the two components oscillating around different average values, but remaining in phase. Notably, the minority component (spin-down for $\px > 0$,
spin-up for $\px < 0$) exhibits a larger oscillation amplitude than the majority one. Finally, for $\px = 0.99$ [Fig.~\ref{fig:dens_profiles}(c)], both
components display only weak density modulations around their mean values which closely resemble those of the uniform plane-wave phase.

\begin{figure}
\centering
\includegraphics{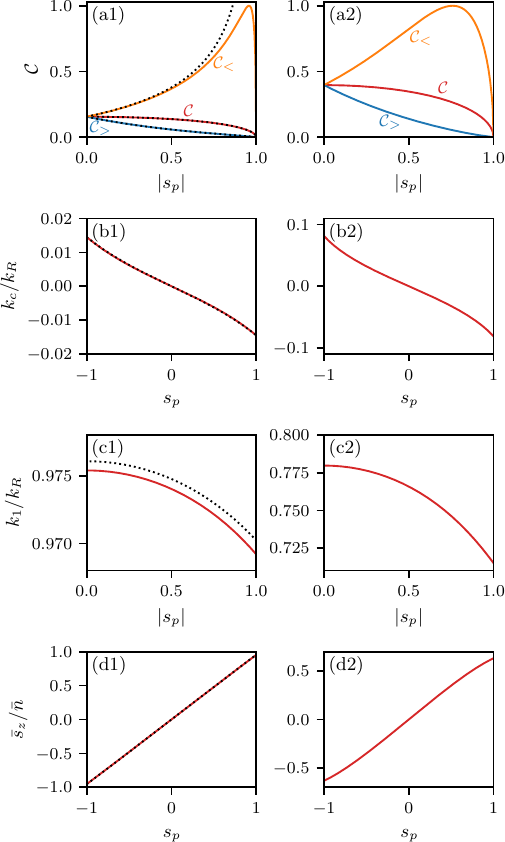}
\caption{The fringe contrast [(a1)-(a2)], the mean condensation wave vector [(b1)-(b2)], the half stripe wave vector [(c1)-(c2)], and the bare
spin polarization density [(d1)-(d2)] as functions of $\px$. In panels [(a1)-(a2)], we show the fringe contrast of the total density (red), as well
as the contrast of the majority (blue) and minority (yellow) bare spin components. Each observable is plotted for two different values of the
Raman coupling: $\hbar\Omega_R / E_R = 1.0$ (left panels) and $\hbar\Omega_R / E_R = 2.6$ (right panels). Solid lines correspond to the numerical
results. In the left column, close to each of these solid lines, we add a black dotted line showing the predictions of the perturbative approach.
The interaction parameters are the same as in Fig.~\ref{fig:dens_profiles}: $G_{dd} / E_R = 1.2$ and $G_{ss} / E_R = 0.32$.}
\label{fig:observables}
\end{figure}

To gain a deeper understanding of the density profiles, it is useful to explicitly evaluate their contrast. For the total density $n$, the contrast
is defined as
\begin{equation}
\mathcal{C} = \frac{n_{\mathrm{max}} - n_{\mathrm{min}}}{n_{\mathrm{max}} + n_{\mathrm{min}}} \, ,
\label{eq:res_contrast_def}
\end{equation}
where $n_{\mathrm{max}}$ and $n_{\mathrm{min}}$ denote the maximum and minimum values of $n$ within one oscillation period, respectively. Generally,
the contrast is an odd function of the Raman coupling $\Omega_R$ since changing the sign of $\Omega_R$ swaps the positions of the density maxima and
minima. Expressing the contrast in terms of the coefficients of the Fourier expansion~\eqref{eq:res_tot_dens} gives
\begin{equation}
\mathcal{C} = \frac{\left| \sum_{\bar{m} = 1}^{+ \infty} \tilde{n}_{2 \bar{m} - 1} \right|}{\bar{n} + \sum_{\bar{m} = 1}^{+ \infty} \tilde{n}_{2\bar{m}}} \, .
\label{eq:res_contrast_coeff}
\end{equation}
At leading order, the contrast depends linearly on $\Omega_R$ and matches (up to a sign) the expression given in Eq.~\eqref{eq:res_dens_coeff_1}.
Figures~\ref{fig:observables}(a1)-(a2) show $\mathcal{C}$ as a function of $\px$ for fixed $\Omega_R$ and interaction parameters. The contrast reaches
its maximum at $\px = 0$, where it coincides with the value reported in Ref.~\cite{Martone2021b}. As $\px$ increases, the contrast decreases, indicating
that the density modulations weaken as the stripes move. The contrast vanishes at $\px = \pm 1$, where the fringes disappear entirely and the system
becomes a uniform plane-wave condensate. Comparison of the two panels reveals that, similarly to the stationary case, the contrast grows with increasing
Raman coupling even at finite $\px$.

The contrast of the density profiles of the individual bare spin components is defined analogously to Eq.~\eqref{eq:res_contrast_def}. As shown in
Figs.~\ref{fig:observables}(a1)-(a2), the contrast $\mathcal{C}_>$ of the majority component, like that of the total density, is maximal at $\px = 0$,
decreases monotonically with increasing $|\px|$, and vanishes at $|\px| = 1$. In contrast, the behavior of the minority component contrast $\mathcal{C}_<$
is nonmonotonic. It initially increases with $|\px|$, reaching the maximum value of $1$ (sooner for larger Raman coupling), and then rapidly drops to zero
as $|\px| \to 1$. At first order in the Raman coupling, $\mathcal{C}_>$ and $\mathcal{C}_<$ can be directly expressed in terms of the coefficients of the
expansions~\eqref{eq:res_tot_dens} and~\eqref{eq:res_spin_dens}. For $\px > 0$, where the majority (minority) component corresponds to spin-up (spin-down),
one finds $\mathcal{C}_> = - (\tilde{n}_1 + \tilde{s}_{z,1}) / [\bar{n}(1 + \px)]$ and $\mathcal{C}_< = - (\tilde{n}_1 - \tilde{s}_{z,1}) / [\bar{n}(1 - \px)]$.
For $\px < 0$, the roles of the two components are interchanged. From Figs.~\ref{fig:observables}(a1)-(a2) we observe that the perturbative prediction for
$\mathcal{C}_>$ matches the numerical results very accurately over the whole range of $|\px|$. However for $\mathcal{C}_<$ the perturbative expression is
reliable only at small $|\px|$. It eventually diverges unphysically as $|\px| \to 1$. This divergence arises because in the limit $|\px| \to 1$ and for
vanishing Raman coupling the denominator in the definition of the contrast [Eq.~\eqref{eq:res_contrast_def}] tends to zero for the minority component.
Consequently, the perturbative approach becomes invalid for calculating $\mathcal{C}_<$ at large $|\px|$. We point out that, since the density fringes
vanish in both bare spin components at $\px = \pm 1$, the system no longer meets the criteria for supersolidity at these points. However, the superfluid
fraction remains finite, although smaller than unity~\cite{Zhang2016,Chen2018,Martone2021b}, indicating that superfluid behavior persists in these limits.

The optimal lengths of the mean condensation wave vector, $k_c$, and of the half stripe wave vector, $k_1$, also exhibit a dependence on $\px$, albeit
much more gradually than the contrast does [see Figs.~\ref{fig:observables}(b1)-(b2) and~\ref{fig:observables}(c1)-(c2)]. These quantities are primarily
governed by single-particle physics, which predicts zero $k_c$ and $k_1$ equal to the wave vector $k_1^{\mathrm{SP}}$ characterizing the single-particle
energy minima. However, interparticle interactions cause $k_c$ to acquire a small but finite value and $k_1$ to slightly deviate from
$k_1^{\mathrm{SP}}$~\cite{Li2012a,Martone2021b}. Following the procedure illustrated in Appendix~\ref{sec:pert_method} one obtains the perturbative formulas
\begin{equation}
k_c = - k_R \px \frac{4 E_R^3 N_{k_c}}{D_1^2} \left( \frac{\hbar\Omega_R}{4 E_R} \right)^2
\label{eq:res_kc}
\end{equation}
and
\begin{equation}
k_1 = k_R \left[ 1 - \frac{2 E_R^2 N_{k_1}}{D_1^2} \left( \frac{\hbar\Omega_R}{4 E_R} \right)^2 \right] \, ,
\label{eq:res_k1}
\end{equation}
where the numerators $N_{k_c}$ and $N_{k_1}$ are given in Eqs.~\eqref{eq:pert_form_N_kc} and~\eqref{eq:pert_form_N_k1}, respectively. One can see that $k_c$
is an odd function of $\px$, and takes negative (positive) values for $\px > 0$ ($\px < 0$). On the other hand, $k_1$ is an even function of $\px$ and
positive by definition. It is maximum at $\px = 0$, where it recovers its value in the stationary stripe phase~\cite{Martone2021b}, and minimum at $\px =
\pm 1$. We have checked that in the limit $\px \to 1$ the combination $k_c + k_1$ tends to the condensation wave vector $k_1^{\mathrm{PW}}$ characterizing
the plane-wave phase, and given by Eq.~\eqref{eq:soc_k1_pw}; similarly, one has $k_c - k_1 \to - k_1^{\mathrm{PW}}$ as $\px \to -1$. Notably, the sum of the
perturbative results~\eqref{eq:res_kc} and~\eqref{eq:res_k1} when evaluated at $\px = 1$ exactly matches the second order expansion of $k_1^{\mathrm{PW}}$
in $\hbar\Omega_R / 4 E_R$. Finally, we emphasize that $k_c$ and $k_1$, like the other observables discussed below, should remain invariant under the sign
change of the Raman coupling. They are therefore even functions of $\Omega_R$.

As already illustrated in Fig.~\ref{fig:dens_profiles}, moving stripes exhibit a population imbalance between the spin-up and spin-down components, resulting
in a finite bare spin polarization. In general, the average bare spin polarization density $\bar{s}_z$ is an odd function of $\px$, being positive for $\px > 0$
and negative for $\px < 0$. Within second order perturbation theory in the Raman coupling, it takes the form
\begin{equation}
\bar{s}_z =
\bar{n} \px \left[ 1 - \frac{2 E_R^2 (2 E_R+G_{ss}) N_{\bar{s}_z}}{D_1^2} \left( \frac{\hbar\Omega_R}{4 E_R} \right)^2 \right] \, ,
\label{eq:res_polar}
\end{equation}
where the coefficient $N_{\bar{s}_z}$ is defined in Eq.~\eqref{eq:pert_form_N_sz}. This expression shows that $\bar{s}_z$ increases linearly with $\px$ for small
$|\px|$, in agreement with the numerical results plotted in Figs.~\ref{fig:observables}(d1)-(d2). The average bare spin polarization density reaches its maximum
magnitude at $\px = \pm 1$, where it coincides with the value $\bar{n} k_1^{\mathrm{PW}} / k_R$ characteristic of the plane-wave phase (see Sec.~\ref{sec:model}).
It is worth noting that $\bar{s}_z$, like the stripe wave vector $k_1$, decreases as the Raman coupling increases. This behavior reflects the reduced separation
between the single-particle energy minima at stronger Raman couplings, which in turn leads to a diminished bare spin polarization associated with each minimum.
We recall that the average momentum density $\bar{p}_x$ is proportional to $\bar{s}_z$ [see Eq.~\eqref{eq:soc_stripe_mom_spin}] and therefore shares the same
properties.

\begin{figure}
\centering
\includegraphics{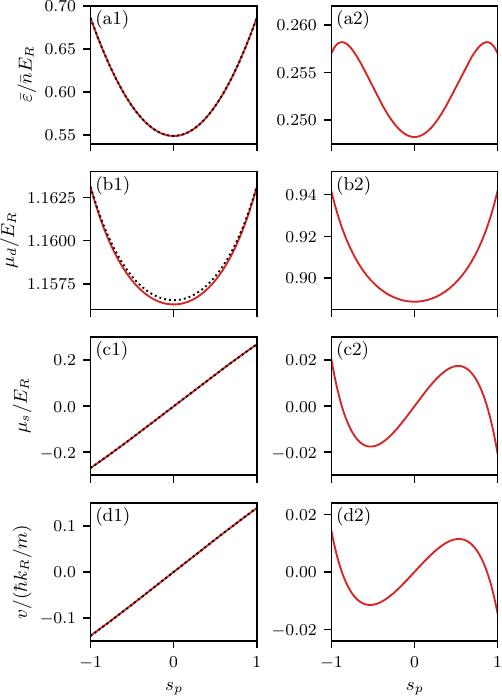}
\caption{The energy per particle [(a1)-(a2)], the mean chemical potential [(b1)-(b2)], the chemical potential semi-difference [(c1)-(c2)], and the fringe velocity
[(d1)-(d2)] as functions of $\px$. As in Fig.~\ref{fig:observables}, each quantity is shown for two different values of the Raman coupling: $\hbar\Omega_R / E_R = 1.0$
(left panels) and $\hbar\Omega_R / E_R = 2.6$ (right panels). Solid lines represent the results of numerical calculations. In the left panels, close to each of these
lines, we add a black dotted line indicating the perturbative predictions. The interaction parameters are the same as in the previous figures: $G_{dd} / E_R = 1.2$
and $G_{ss} / E_R = 0.32$.}
\label{fig:energies}
\end{figure}

The perturbative approach also allows for the derivation of analytical expressions for key thermodynamic quantities. The average energy density, obtained from
Eq.~\eqref{eq:soc_en}, reads
\begin{equation}
\bar{\varepsilon}
= \bar{n} \left[\frac{G_{dd}}{2} + \frac{G_{ss}}{2} \px^2 - \frac{E_R^2 N_{\bar{\varepsilon}}}{D_1} \left( \frac{\hbar\Omega_R}{4 E_R} \right)^2\right] \, ,
\label{eq:res_en}
\end{equation}
while the mean chemical potential is given by
\begin{equation}
\mu_d = G_{dd} + \frac{2 E_R^2 N_{\mu_d}}{D_1^2} \left( \frac{\hbar\Omega_R}{4 E_R} \right)^2 \, ,
\label{eq:res_mu_d}
\end{equation}
and the chemical potential semi-difference takes the form
\begin{equation}
\mu_s = \px \left[ G_{ss} - \frac{2 E_R^2 (2 E_R+G_{ss}) N_{\mu_s}}{D_1^2} \left( \frac{\hbar\Omega_R}{4 E_R} \right)^2 \right] \, .
\label{eq:res_mu_s}
\end{equation}
The coefficients $N_{\bar{\varepsilon}}$, $N_{\mu_d}$, and $N_{\mu_s}$ are given in Eqs.~\eqref{eq:pert_form_N_eps}, \eqref{eq:pert_form_N_mu_d},
and~\eqref{eq:pert_form_N_mu_s}, respectively. These expressions satisfy the thermodynamic relations~\eqref{eqs:soc_en_dens_der}. Using the identity
$v = \mu_s / \hbar k_1$, and combining Eqs.~\eqref{eq:res_k1} and~\eqref{eq:res_mu_s}, one obtains the perturbative expression for the fringe velocity,
\begin{equation}
v = \frac{\px}{\hbar k_R} \left[ G_{ss} - \frac{8 E_R^4 N_v}{D_1^2} \left( \frac{\hbar\Omega_R}{4 E_R} \right)^2 \right] \, ,
\label{eq:res_vel}
\end{equation}
with $N_v$ defined in Eq.~\eqref{eq:pert_form_N_v}. Both $\bar{\varepsilon}$ and $\mu_d$ are even functions of $\px$, reaching their minimum at $\px = 0$,
where they decrease to the values derived in Ref.~\cite{Martone2021b}. At small $\px$, they increase quadratically [see Figs.~\ref{fig:energies}(a1)-(a2)
and~\ref{fig:energies}(b1)-(b2)]. In contrast, both $\mu_s$ and $v$ are odd functions of $\px$, and grow linearly with positive slope for small $|\px|$,
as shown in Figs.~\ref{fig:energies}(c1)-(c2) and~\ref{fig:energies}(d1)-(d2). In the limit $\px \to \pm 1$, the average energy density approaches that of
the plane-wave phase~\cite{Li2012a,Martone2012}. The behavior of $\bar{\varepsilon}$ as a function of $\px$ depends strongly on the Raman coupling strength.
At low $\Omega_R$, the quantity $\bar{\varepsilon}$ increases monotonically with $|\px|$ and attains its maximum at $\px = \pm 1$, as seen in
Fig.~\ref{fig:energies}(a1). In contrast, at large $\Omega_R$, the quantity $\bar{\varepsilon}$ develops a global maximum at an intermediate value of $|\px|$
and decreases thereafter, reaching local minima at $\px = \pm 1$ corresponding to the plane-wave states [see Fig.~\ref{fig:energies}(a2)]. Due to the thermodynamic
relation in Eq.~\eqref{eq:soc_en_dens_der_s}, this qualitative change also affects the behavior of $\mu_s$ and $v$. While the mean chemical potential $\mu_d$
remains a convex function of $\px$, $\mu_s$ at large $\Omega_R$ develops two stationary extrema, a minimum at negative $\px$ and a maximum at positive
$\px$, corresponding to inflection points of the average energy density. Both $\mu_s$ and $v$ vanish at the local maxima of $\bar{\varepsilon}$.
Interestingly, in the large-$|\px|$ regime, the fringe velocity may acquire the sign opposite to that of the dressed spin imbalance. However, this regime is
dynamically unstable, as discussed in Sec.~\ref{subsec:bogo_sound}. Finally, within the metastable window between the critical point $\Omega_{\mathrm{cr1}}$ and
the spinodal point $\Omega_{\mathrm{sp}}$ the stationary stripe configuration at $\px = 0$ represents only a local minimum of the energy per particle, while the
plane-wave states at $\px = \pm 1$ are energetically favored. Note that, although $\mu_d$ and $\mu_s$ become ill defined individually at $\px = \pm 1$, the
combinations $\mu_d + \mu_s$ (for $\px \to 1$) and $\mu_d - \mu_s$ (for $\px \to -1$) smoothly approach the chemical potential of the plane-wave
phase~\cite{Martone2012}.

\section{Excitation spectrum of moving stripe patterns}
\label{sec:exc_spectrum}
In this section, we examine dynamical properties of moving stripe patterns. We begin by outlining the Bogoliubov theory for studying small oscillations
around the stripe background in the comoving frame, where the fringes remain stationary (Sec.~\ref{subsec:bogo_theory}). We then compute the Bogoliubov spectrum
and the associated sound velocities, highlighting the emergence of both energetic and dynamical instabilities at large dressed spin imbalance
(Sec.~\ref{subsec:bogo_sound}).

\subsection{Bogoliubov theory in the comoving frame}
\label{subsec:bogo_theory}
The analysis in the previous sections was carried out in the laboratory frame, that is, the rest frame of the Raman lasers responsible for generating the
spin-orbit coupling. An equivalent description can be formulated in the comoving frame, which corresponds to the rest frame of the crystal pattern. The space
and time coordinates in the two frames are related by the usual transformations $x' = x - v t$, $y' = y$, $z' = z$, and $t' = t$ (throughout this work,
primed quantities refer to the comoving frame). We have already seen that the order parameter of a moving stripe pattern becomes stationary when expressed
in terms of $\vec{r}' = (x',y',z')$ and $t'$. Consequently, the comoving frame provides the most suitable framework in which to study the dynamical properties
of traveling stripes, as their excitation spectrum is well defined in this frame.

The order parameter $\Psi'(\vec{r}',t')$ in the comoving frame satisfies the time-dependent Gross-Pitaevskii equation
\begin{equation}
\mi \hbar \partial_{t'} \Psi' = h_{\mathrm{SO}}' \Psi' + g_{dd} \left(\Psi'^\dagger \Psi'\right) \Psi'
+ g_{ss} \left(\Psi'^\dagger \sigma_z \Psi'\right) \sigma_z \Psi' \, .
\label{eq:exc_td_gp_comov}
\end{equation}
This equation differs from its counterpart in the laboratory frame, Eq.~\eqref{eq:soc_td_gp}, due to an additional detuning term arising from the
transformed spin-orbit Hamiltonian,
\begin{equation}
h_{\mathrm{SO}}' = h_{\mathrm{SO}} - v \hbar k_R \sigma_z \, .
\label{eq:exc_soc_ham_transf}
\end{equation}
This extra term can be understood by noting that in the comoving frame the Raman lasers move with velocity $-v$ along $x$, see Fig.~\ref{fig:scenarios}(b2).
As a result, the frequency of the light field experienced by the atoms undergoes a Doppler shift of $- 2 v k_R$, which leads to the corresponding shift in the
Raman detuning $\delta_R$ by the same amount~\cite{Zheng2013,Hamner2015}.

In order to apply the Bogoliubov theory~\cite{Pitaevskii_Stringari_book,Pethick_Smith_book,Castin_review} and study small oscillations on top of a
moving stripe pattern, we write the order parameter in the comoving frame in the form
\begin{equation}
\Psi'(\vec{r}',t') = \me^{-\mi \mu' t' / \hbar} \me^{- \mi m v x' / \hbar} \left[ \Psi_0(x') + \delta \Psi(\vec{r}',t') \right] \, ,
\label{eq:exc_wf}
\end{equation}
with $\mu' = \mu + m v^2/2$. Substituting this expression into Eq.~\eqref{eq:exc_td_gp_comov}, and neglecting for a moment the small fluctuation term
$\delta \Psi$, one recovers the time-independent equation~\eqref{eq:soc_ti_gp_p} obeyed by $\Psi_0$. This confirms the consistency of the descriptions
in the laboratory and comoving frames. The presence of the $\me^{- \mi m v x' / \hbar}$ factor in Eq.~\eqref{eq:exc_wf} indicates that in the comoving
frame superflow occurs with velocity $-v$ along the $x$ direction, as shown in Fig.~\ref{fig:scenarios}(b2).

The linear terms in $\delta\Psi$ that emerge after inserting the Ansatz~\eqref{eq:exc_wf} into the Gross-Pitaevskii equation~\eqref{eq:exc_td_gp_comov}
yield an evolution equation for the small fluctuation,
\begin{equation}
\mi\hbar\partial_{t'} \delta \Psi
= \left( h_{\mathrm{SO}}
- \mu - v p_x + h_{\mathrm{D}} \right) \delta \Psi
+ h_{\mathrm{C}} \delta \Psi^* \, ,
\label{eq:exc_eq_Psi}
\end{equation}
where
\begin{subequations}
\label{eq:exc_hD_hC}
\begin{align}
\begin{split}
h_{\mathrm{D}} =
{}&{} g_{dd} (\Psi_0^\dagger\Psi_0 \mathbb{I}_2 + \Psi_0 \Psi_0^\dagger) \\
&{} + g_{ss} \left[ (\Psi_0^\dagger \sigma_z \Psi_0) \sigma_z
+ (\sigma_z\Psi_0) (\sigma_z\Psi_0)^\dagger \right] \, ,
\end{split}
\label{eq:exc_hD} \\
h_{\mathrm{C}} =
{}&{} g_{dd} \Psi_0 \Psi_0^T
+ g_{ss} (\sigma_z\Psi_0) (\sigma_z\Psi_0)^T \, ,
\label{eq:exc_hC}
\end{align}
\end{subequations}
and $\mathbb{I}_2$ is the $2 \times 2$ identity matrix. As is customary, we look for solutions of the linearized problem~\eqref{eq:exc_eq_Psi} with a single,
possibly complex, oscillation frequency:
\begin{equation}
\delta \Psi(\vec{r}',t') = U_{b,\vec{k}}(\vec{r}') \me^{- \mi \omega_{b,\vec{k}} t'} + V_{b,\vec{k}}^*(\vec{r}') \me^{\mi \omega_{b,\vec{k}}^* t'} \, .
\label{eq:exc_lin_wf}
\end{equation}
The two-component amplitudes $U_{b,\vec{k}}$ and $V_{b,\vec{k}}$ depend on position and are defined up to a global complex normalization factor. The meaning of
the indices $b$ and $\vec{k}$ will be clarified shortly. Substituting Eq.~\eqref{eq:exc_lin_wf} into Eq.~\eqref{eq:exc_eq_Psi} and collecting terms oscillating
at the same Bogoliubov frequency $\omega_{b,\vec{k}}$ we obtain an eigenvalue equation for $\omega_{b,\vec{k}}$ and the corresponding amplitudes:
\begin{equation}
(\eta \mathcal{B} - v p_x)
\begin{pmatrix}
U_{b,\vec{k}} \\
V_{b,\vec{k}}
\end{pmatrix}
= \hbar \omega_{b,\vec{k}}
\begin{pmatrix}
U_{b,\vec{k}} \\
V_{b,\vec{k}}
\end{pmatrix} \, ,
\label{eq:exc_eq_UV}
\end{equation}
where $\eta = \operatorname{diag}(1,1,-1,-1)$ and
\begin{equation}
\mathcal{B} =
\begin{pmatrix}
h_{\mathrm{SO}} - \mu + h_{\mathrm{D}} & h_{\mathrm{C}} \\
h_{\mathrm{C}}^* & (h_{\mathrm{SO}} - \mu + h_{\mathrm{D}})^*
\end{pmatrix} \, .
\label{eq:exc_B_mat}
\end{equation}
These expressions formally coincide with those derived for the stationary stripe phase~\cite{Li2013,Martone2021b}, except for the additional Doppler term
$- v p_x$ on the left hand side of Eq.~\eqref{eq:exc_eq_UV}. As in the $v = 0$ case, we seek solutions in the form of Bloch waves:
\begin{equation}
\begin{pmatrix}
U_{b,\vec{k}}(\vec{r}') \\
V_{b,\vec{k}}(\vec{r}')
\end{pmatrix}
=
\frac{\me^{\mi (\vec{k} + \eta \vec{k}_c) \cdot \vec{r}'}}{\sqrt{V}}
\sum_{\text{$\bar{m}$ odd}}
\begin{pmatrix}
\tilde{U}_{b,\vec{k},\bar{m}} \\
\tilde{V}_{b,\vec{k},\bar{m}}
\end{pmatrix}
\me^{\mi \bar{m} k_1 x'} \, .
\label{eq:exc_UV_exp}
\end{equation}
Here, $\vec{k}$ is the excitation quasimomentum, while the $\tilde{U}_{b,\vec{k},\bar{m}}$'s and $\tilde{V}_{b,\vec{k},\bar{m}}$'s are two-component expansion
coefficients. Combining Eqs.~\eqref{eq:exc_eq_UV} and~\eqref{eq:exc_UV_exp} we obtain an infinite-dimensional eigenvalue problem for these coefficients and the
corresponding Bogoliubov frequencies $\omega_{b,\vec{k}}$. A detailed discussion of the case $v = 0$ can be found in Ref.~\cite{Martone2018}. At fixed $\vec{k}$,
one obtains an infinite set of eigenfrequencies that form a band structure as $\vec{k}$ varies~\cite{Li2013,Martone2021b}. This justifies the use of an
additional subscript, the band index $b$, to label the amplitudes and the frequency of each Bogoliubov mode. The most general solution to
Eq.~\eqref{eq:exc_eq_Psi} can then be expressed as a sum over all physically distinct excitation bands and quasimomenta.

The Bogoliubov spectrum obtained by solving the eigenvalue equation~\eqref{eq:exc_eq_UV} exhibits several notable properties:
\begin{enumerate}
\renewcommand{\labelenumi}{\arabic{enumi})}
\item complex frequencies occur in complex conjugate pairs. This follows from the fact that the operator $\eta \mathcal{B} - v p_x$ differs from its
Hermitian adjoint by a unitary transformation: $\eta (\eta \mathcal{B} - v p_x) \eta^{-1} = (\eta \mathcal{B} - v p_x)^\dagger$~\cite{Castin_review};
\item each mode has a conjugate counterpart that describes the same physical oscillation. Specifically, if the pair $(U_{b,\vec{k}},V_{b,\vec{k}})$ solves
Eq.~\eqref{eq:exc_eq_UV} with frequency $\omega_{b,\vec{k}}$, then the pair $(V_{b,\vec{k}}^*,U_{b,\vec{k}}^*)$ is a solution with frequency
$- \omega_{b,\vec{k}}^*$~\cite{Castin_review}. However, because of the structure of the Ansatz~\eqref{eq:exc_lin_wf}, these two solutions describe the same
physical excitation;
\item the spectrum is periodic in the $x$-component of the quasimomentum. Specifically, for each band $b$, the Bogoliubov frequency satisfies the periodicity
condition $\omega_{b,\vec{k} + 2 k_1 \unitvecx} = \omega_{b,\vec{k}}$, with period $2 k_1$ corresponding to the extent of the first Brillouin zone.
\end{enumerate}
The first property implies that a moving stripe pattern is dynamically stable as long as its excitation spectrum is entirely real. When this condition is met,
the second property ensures that there are equal numbers of solutions of Eq.~\eqref{eq:exc_eq_UV} with positive and negative frequency. These solutions
obey the orthonormalization relation~\cite{Castin_review}
\begin{equation}
\begin{split}
&{} \int_V \dif^3 r' \, \left[ U_{b',\vec{k}'}^\dagger(\vec{r}') U_{b,\vec{k}}(\vec{r}') - V_{b',\vec{k}'}^\dagger(\vec{r}') V_{b,\vec{k}}(\vec{r}') \right] \\
&{} = \mathcal{N}_{b,\vec{k}} \delta_{b b'} \delta_{\vec{k} \vec{k'}} \, ,
\end{split}
\label{eq:exc_UV_orthonorm}
\end{equation}
which has two key implications. First, it defines orthogonality between modes belonging to different bands ($b' \neq b$) or with different quasimomenta ($\vec{k}'
\neq \vec{k}$). Second, in the case $b' = b$ and $\vec{k}' = \vec{k}$, it provides the norm $\mathcal{N}_{b,\vec{k}}$ of a given Bogoliubov mode. Importantly,
each physical oscillation corresponds to a pair of solutions with opposite norms. In plotting the Bogoliubov spectrum, such as in Fig.~\ref{fig:Bogo_spectrum}, it
is customary to include only the frequencies of the positive norm solutions. Furthermore, the system is said to be energetically stable if all positive norm
solutions have positive frequency. In the next section, we will show that moving stripe patterns can lose energetic or dynamical stability at sufficiently large
dressed spin imbalance. Finally, due to the periodicity of the spectrum, it suffices to consider values of the $x$ component of $\vec{k}$ within the first
Brillouin zone, that is, $- k_1 \leq k_x < k_1$.\footnote{Here we depart from the conventions of most previous literature~\cite{Li2013,Chen2018,Martone2018,
Martone2021b,Xia2023,Lyu2024}, which takes the first Brillouin zone for the excitation quasimomentum as the interval $0 \leq k_x < 2 k_1$.} In contrast, the $y$
and $z$ components are unconstrained.

\subsection{Bogoliubov spectrum and sound velocities}
\label{subsec:bogo_sound}
To numerically compute the Bogoliubov spectrum, we truncate the expansions~\eqref{eq:exc_UV_exp} by retaining only a finite number of terms. Specifically,
we include components with $- 9 \leq \bar{m} \leq 9$, following the approach used in Sec.~\ref{subsec:eval_ord_param} for evaluating $\Psi_0$. This truncation
reduces Eq.~\eqref{eq:exc_eq_UV} to a finite-dimensional eigenvalue problem, which can be readily solved. A detailed discussion of the computational procedure
is given in Ref.~\cite{Martone2018}. In Fig.~\ref{fig:Bogo_spectrum}, we display the lowest-lying bands of the computed spectrum for excitation quasimomentum
$\vec{k}$ aligned along the $x$ axis. The figure consists of two rows of panels, corresponding to the two values of the Raman coupling considered in this work.
Within each row, we vary the value of $\px$ to illustrate four representative cases:
\begin{figure*}[htb]
\centering
\includegraphics[scale=0.75]{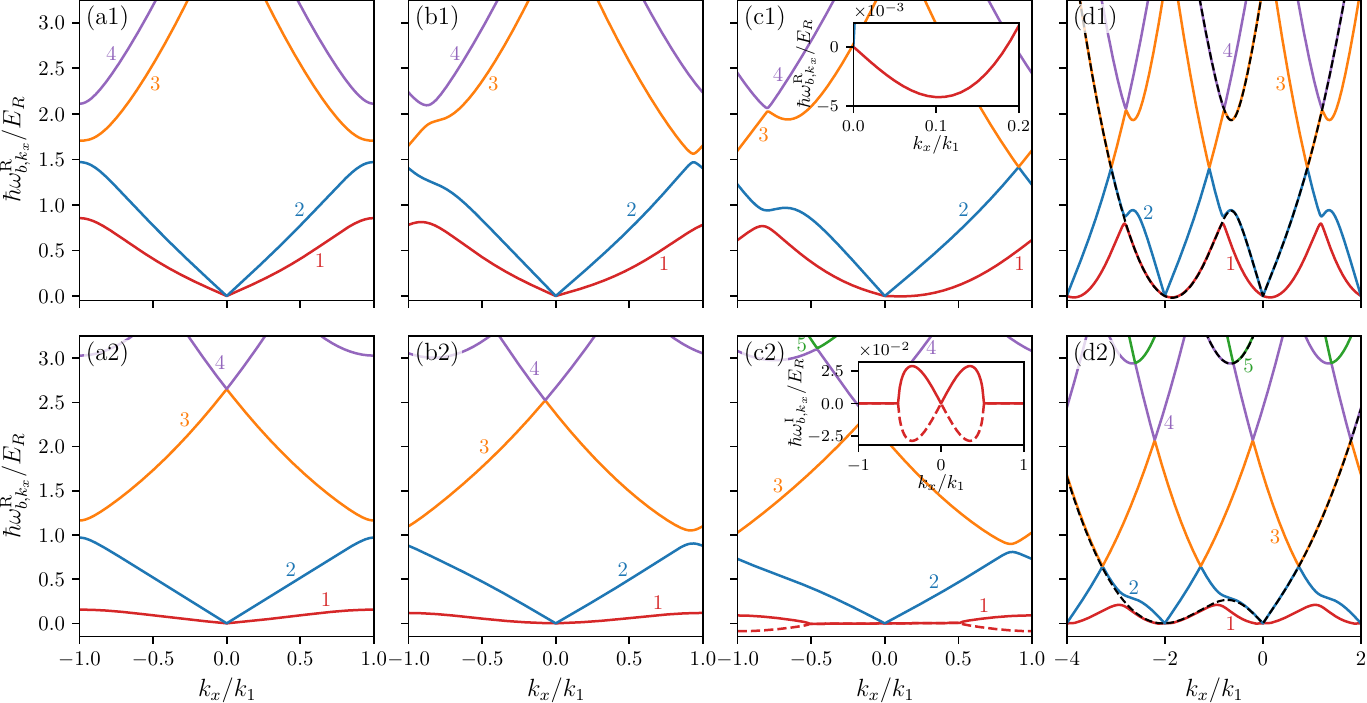}
\caption{Excitation spectrum of a spin-orbit-coupled BEC in the stripe phase for various values of $\px$: (a1)-(a2) $\px = 0$ (stationary stripes), (b1)-(b2)
$\px = 0.5$, (c1) $\px = 0.95$, (c2) $\px = 0.75$, and (d1)-(d2) $\px = 0.99$. The top-row panels correspond to $\hbar\Omega_R / E_R = 1.0$, while those in
the bottom row correspond to $\hbar\Omega_R / E_R = 2.6$. The excitation quasimomentum is taken along the $x$ direction. Only the lowest-lying excitation bands
are shown; the band index $b$ is indicated next to each curve. In (c1), the inset shows a magnified view in the vicinity of $k_x = 0$ on the positive-$k_x$ side,
where the lowest gapless band exhibits a negative frequency. In all panels, we plot only the real part of the frequency, $\omega_{b,k_x}^{\mathrm{R}}$; the
imaginary part, $\omega_{b,k_x}^{\mathrm{I}}$, is zero except for the lowest gapless band in (c2) (see inset) and in (d2), where it is nonzero but very small
and omitted for clarity. The red dashed line in (c2) displays the solution $\omega_{1,k_x}^*$ in the dynamically unstable region and $- \omega_{1,-k_x}$
in the dynamically stable region, both obtained from Eq.~\eqref{eq:exc_eq_UV}. In panels (d1)-(d2), the black dashed curves represent the two branches of the
excitation spectrum in the plane-wave phase, computed using the same parameters as in the case of the stripe-phase spectrum, and with an added Doppler shift
(see text for details). The interaction parameters are as in previous figures: $G_{dd} / E_R = 1.2$ and $G_{ss} / E_R = 0.32$.}
\label{fig:Bogo_spectrum}
\end{figure*}
\begin{enumerate}
\renewcommand{\labelenumi}{(\alph{enumi})}
\item Spectrum of stationary stripes ($\px = 0$), previously computed in Refs.~\cite{Li2013,Martone2021b}. The two lowest bands are gapless and exhibit
linear dispersion near $k_x = 0$, i.e., the center of the first Brillouin zone. The two zero-frequency modes at $k_x = 0$ correspond to the Goldstone
modes arising from the spontaneous breaking of global-phase and translation symmetries. Recall that, in the regime $g_{dd} > g_{ss}$ considered here, the
upper (lower) phonon branch has predominantly density (spin) character~\cite{Li2013}, corresponding to the superfluid (crystalline) nature of the stripe
phase~\cite{Martone2021b}. The roles are interchanged when $g_{ss} > g_{dd}$.
\item Spectrum of moving stripes at intermediate dressed spin imbalance ($\px = 0.5$). Most features remain similar to the stationary case. However, the
bands no longer exhibit reflection symmetry with respect to the center of the Brillouin zone. This is due to the fact that the transformation $k_x \to - k_x$
transforms the spectrum of a stripe pattern to the one with opposite $\px$. Additionally, the gaps between bands are narrower for moving stripes, and this
trend becomes more pronounced as $|\px|$ increases.
\item Spectrum in the instability regime ($\px = 0.95$ for the top row, $\px = 0.75$ for the bottom row). The behavior depends on the Raman coupling. At
lower Raman coupling (top row), a portion of the lowest gapless band dips into negative frequency, particularly right of $k_x = 0$ (or left of $k_x = 0$
for negative $\px$), as shown in the inset of panel (c1). This signals the onset of energetic instability for moving stripes at large dressed spin imbalance.
Physically, this means that in these configurations the increase in energy cost associated with the large bare spin imbalance [see Fig.~\ref{fig:observables}(d1)]
is no longer sufficiently compensated by the energy decrease resulting from the reduction of the stripe contrast [Fig.~\ref{fig:observables}(a1)]. Conversely,
at higher Raman coupling (bottom row), the system first exhibits energetic instability (now left of $k_x = 0$), followed by the emergence of a nonzero
imaginary part in the lowest band near $k_x = 0$, as seen in panel (c2). According to the discussion in the previous section, this indicates that moving
stripe patterns become dynamically unstable at high Raman coupling and high dressed spin imbalance. We speculate that this dynamical instability would
lead to a spatial separation of the two dressed spin components of the stripe order parameter, resulting in the formation of multiple plane-wave domains.
\item Spectrum near the melting limit ($\px = 0.99$). Unlike the previous panels, which focus on the first Brillouin zone, here we plot the spectrum over
the range $- 4 k_1 \leq k_x < 2 k_1$, whose length is three times that of the first Brillouin zone. One observes a significant narrowing of the band gaps,
heralding their complete closure at the stripe melting point. At this point, segments from different bands and Brillouin zones merge to form the dispersion
relations $\omega_{\pm,\vec{k}} - v k_x$ (dashed lines), where $\omega_{\pm,\vec{k}}$ denote the lower ($-$) and upper ($+$) branches of the excitation
spectrum in the plane-wave phase~\cite{Martone2012}.\footnote{Figure~\ref{fig:Bogo_spectrum}(d1)-(d2) may suggest that the Bogoliubov frequencies remain
periodic in $k_x$ even in the plane-wave phase. However, one must recall that $\vec{k}$ represents the excitation \textit{quasimomentum}. The spectrum
loses periodicity when plotted as a function of the excitation \textit{momentum}, which is well defined only for $|\px| = 1$; see the related discussion in
Ref.~\cite{Martone2021b}.} The Doppler shift $- v k_x$ appears because the spectrum is evaluated in the comoving frame. Notably, the density phonon modes
(upper gapless branch) of the stripe phase smoothly evolve into the single phonon branch of the plane-wave phase. More intriguingly, the spin phonon modes
(lower gapless branch) transform into the well-known roton minimum of the plane-wave excitation spectrum near $k_x = - 2 k_1^{\mathrm{PW}}$ (or $k_x =
+ 2 k_1^{\mathrm{PW}}$ for the plane-wave state with negative condensation momentum, corresponding to $\px = - 1$)~\cite{Zheng2012,Martone2012,Zheng2013,
Khamehchi2014,Ji2015}.
\end{enumerate}
The occurrence of an energetic instability in the spectrum of moving stripes is consistent with the behavior observed on the plane-wave side of the phase
diagram. In this regime, the roton gap progressively decreases as the Raman coupling $\Omega_R$ is lowered towards the transition to the stripe
phase~\cite{Martone2012,Khamehchi2014,Ji2015}. Eventually, the gap closes at a spinodal point slightly below the critical value $\Omega_{\mathrm{cr1}}$.
Upon further reduction of $\Omega_R$, the Bogoliubov modes near the roton minimum acquire negative frequency. These effects can be seen in the dashed lines
of Fig.~\ref{fig:Bogo_spectrum}(d1)-(d2), although the plane-wave spectrum is modified in the comoving frame due to the Doppler shift term $- v k_x$.
In contrast, the plane-wave phase remains dynamically stable for all values of $\Omega_R$. This implies that the dynamical instability of moving stripe
patterns must vanish once the limit $\px \to \pm 1$ is reached. Our numerical analysis confirms this expectation: beyond a certain threshold, the dynamically
unstable region begins to shrink as $|\px|$ increases, and it disappears entirely when $|\px| = 1$. This behavior reflects the trend observed in the sound
velocities, as discussed below.

\begin{figure}[htb]
\centering
\includegraphics{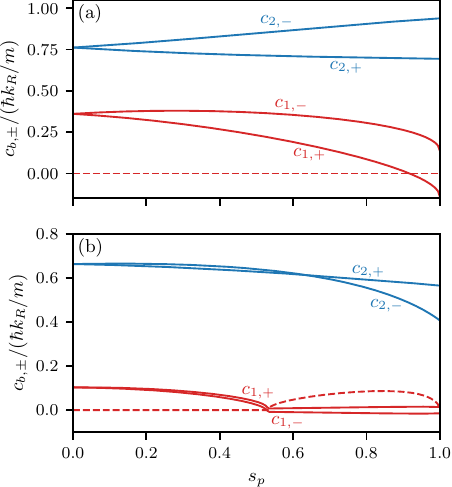}
\caption{Velocities of spin ($b=1$, red lines) and density ($b=2$, blue lines) sound waves as functions of $\px$ for (a) $\hbar\Omega_R / E_R = 1.0$
and (b) $\hbar\Omega_R / E_R = 2.6$. Results for both forward ($c_{b,+}$) and backward ($c_{b,-}$) propagation relative to the $x$ direction are shown.
Owing to the symmetry property $c_{b,+}(-\px) = c_{b,-}(\px)$, we restrict the plot to positive values of $\px$. For the spin sound velocities, we plot
both the real parts (solid lines) and the imaginary parts (dashed lines), with the latter being identical for $c_{1,+}$ and $c_{1,-}$. The interaction
parameters are the same as in previous figures: $G_{dd} / E_R = 1.2$ and $G_{ss} / E_R = 0.32$.}
\label{fig:sound_vel}
\end{figure}

We now analyze in more detail the long wavelength behavior of the two gapless excitation bands ($b = 1,2$), focusing in particular on the velocities of
sound waves propagating parallel ($c_{b,+}$) and antiparallel ($c_{b,-}$) to the $x$ direction. These velocities characterize the linear dispersions
of the gapless bands near the center of the first Brillouin zone, namely, $\omega_{b,k_x} = c_{b,+} k_x + O(k_x^3)$ for $k_x > 0$ and
$\omega_{b,k_x} = - c_{b,-} k_x + O(k_x^3)$ for $k_x < 0$. As illustrated in Fig.~\ref{fig:sound_vel}, for both the spin ($b = 1$) and density ($b = 2$)
sound modes, the velocities in the positive $x$ direction differ from those in the negative $x$ direction, except at $\px = 0$, where the two become
equal. This symmetry at $\px = 0$ reflects the underlying symmetry of the excitation spectrum with respect to the center of the Brillouin zone, as discussed
earlier. More generally, at arbitrary $\px$ one has $c_{b,+}(-\px) = c_{b,-}(\px)$.

The density sound velocities $c_{2,\pm}$ are always real and positive. In the limit $|\px| \to 1$, they approach the expressions $c_{\pm x} \mp v_x$,
where $c_{+x}$ ($c_{-x}$) denotes the speed of density sound waves propagating along (against) the $x$ direction in the plane-wave phase. Note that
$c_{+x} \neq c_{-x}$ in the presence of nonzero spin-dependent interactions ($g_{ss} \neq 0$)~\cite{Martone2012}.

In contrast, the behavior of the spin sound velocities reveals the onset of instability. As $|\px|$ increases, one of the two velocities $c_{1,\pm}$
becomes negative at a certain point, indicating the emergence of an energetic instability, as previously discussed. For high Raman coupling, this
instability is soon followed by the appearance of complex sound velocities: both $c_{1,+}$ and $c_{1,-}$ acquire imaginary parts of equal magnitude.
These imaginary parts first grow with $|\px|$, then decrease, and vanish as $|\px| \to 1$. In this limit, the real parts tend towards opposite values,
satisfying $c_{1,+} = - c_{1,-}$. This behavior confirms that the emergence of energetic instability is a necessary precursor to ensuring continuity
of the excitation spectrum slope in the comoving frame at the points $k_x = \pm 2 k_1$ in the plane-wave phase.

\section{Experimental perspectives}
\label{sec:exp_persp}
To discuss the feasibility of observing moving stripe patterns, we first note that the interaction parameters used in this work differ significantly from
those of the original experiment reported in Ref.~\cite{Lin2011}. The states of ${}^{87}$Rb employed in that experiment are indeed characterized by a very
low degree of miscibility, as evidenced by the small ratio $g_{ss} / g_{dd} \sim 10^{-3}$, resulting in a small critical Raman coupling
$\hbar\Omega_{\mathrm{cr1}} / E_R = 0.19$. This limits the maximum achievable contrast of the stripes and thus restricts the possibility of observing
significant supersolidity effects. Several recent works~\cite{Li2017,Putra2020,Chisholm2024} have implemented strategies to address this limitation,
leading to the detection of the density modulations characterizing the stripe phase. For example, by implementing spin-orbit coupling between two orbital
states within a superlattice potential~\cite{Li2016}, it has been possible to reduce the spatial overlap between the wave functions of the two bare spin
components along a direction parallel to the stripes (a similar effect can also be achieved by applying a quasi-two-dimensional spin-dependent trapping
potential~\cite{Martone2014,Martone2015,deHond2022}). This resulted in a significant increase in the effective value of the ratio $g_{ss} / g_{dd}$, of more
than two orders of magnitude compared to Ref.~\cite{Lin2011}, and led to the first observation of the stripes by Ketterle's group at MIT~\cite{Li2017}.
Alternatively, one can use atomic species with tunable interactions, as demonstrated in~\cite{Chisholm2024} employing ${}^{41}$K atoms. Stripes with high
contrast can also be obtained by rapidly ramping the Raman coupling to some very large value, as done in~\cite{Putra2020}. Although these strategies have
mostly, though not exclusively (see below), been employed to detect stationary stripes, they can also be used to observe traveling stripes, for which
operating at sufficiently large Raman coupling is crucial to compensate for the reduction in contrast due to the stripe motion.

Moving stripe patterns can also be observed in configurations with finite Raman detuning or asymmetric intraspecies couplings. In these cases, stationary
stripes are characterized by a nonzero dressed spin imbalance $\px^{\mathrm{st}}$; traveling stripes then appear for $\px \neq \px^{\mathrm{st}}$. A first
demonstration of this mechanism was provided by the MIT experiment~\cite{Li2017}, which reported the emergence of moving stripes with $\px = 0$ following
the application of a Raman detuning. This generates a chemical potential difference between the two dressed spins even when their populations are equal.
In contrast, the configuration explored in this work involves a population imbalance between the dressed spins, with the resulting stripe motion driven by
spin-dependent interactions. This scenario can be realized experimentally by first preparing the BEC in a superposition of two atomic hyperfine states with
different weights. Raman coupling can then be adiabatically introduced and ramped up to the desired strength. In typical experimental conditions, such as
those of Refs.~\cite{Li2017,Chisholm2024}, the fringe velocity can reach values up to a few millimeters per second, which lies within current experimental
resolution. The traveling density modulations can be probed using Bragg scattering: a moving fringe pattern imparts a phase shift to the diffracted beam,
which can be detected through interferometric techniques.

\section{Conclusions}
\label{sec:concl}
In this work, we have investigated the properties of traveling supersolid patterns in a spin-orbit-coupled Bose-Einstein condensate. The motion of the
density fringes stems from a population imbalance between the two minima of the single-particle dispersion. Antiferromagnetic spin-dependent interactions
create a chemical potential difference which sets the stripe velocity. These moving stripes exhibit asymmetric densities of the two bare spin components,
notable bare spin polarization, reduced contrast, and higher energy than that of stationary stripes. As the dressed spin imbalance increases,
these effects intensify, eventually leading to the disappearance of the fringes and the emergence of a uniform plane-wave condensate. This indicates that
traveling stripes interpolate between the stationary stripe and plane-wave phases. By analyzing the excitation spectrum, we uncover energetic and dynamical
instabilities at high dressed spin imbalance, mainly affecting the spin phonon branch that becomes the mode with the roton minimum in the plane-wave phase.
Our findings extend prior work~\cite{Geier2021,Geier2023} on the slow motion of stripes induced by the release of a weak spin perturbation, where the
propagation of fringes manifests the crystal Goldstone mode of the stripe phase. The parameter regime required to observe traveling stripe patterns is
within reach of current experimental setups, opening new perspectives for the exploration of dynamical aspects of supersolidity in ultracold quantum gases.

\begin{acknowledgments}
Useful discussions and collaborations with K. T. Geier, Ph. Hauke, W. Ketterle, Y. Li, A. Recati, S. Stringari, and D. Trypogeorgos are acknowledged.
This work was supported by the Italian Ministry of University and Research (MUR) through the PNRR MUR project: `National Quantum Science and Technology
Institute' - NQSTI (PE0000023) and the PNRR MUR project: `Integrated Infrastructure Initiative in Photonic and Quantum Sciences' - I-PHOQS (IR0000016).
We acknowledge the support of the Quantum Optical Networks based on Exciton-polaritons - (Q-ONE) funding from the HORIZON-EIC-2022-PATHFINDER CHALLENGES
EU programme under grant agreement No. 101115575, and of the Neuromorphic Polariton Accelerator - (PolArt) funding from the Horizon-EIC-2023-Pathfinder
Open EU programme under grant agreement No. 101130304. Views and opinions expressed are however those of the author(s) only and do not necessarily
reflect those of the European Union or European Innovation Council and SMEs Executive Agency (EISMEA). Neither the European Union nor the granting
authority can be held responsible for them.
\end{acknowledgments}

\appendix

\section{Perturbative method}
\label{sec:pert_method}
This appendix presents a detailed methodology employed to derive the perturbative formulas discussed in the main text. The general procedure described in
Sec.~\ref{subsec:pert_gen} closely follows the approach of Ref.~\cite{Martone2021b}, with appropriate generalizations to incorporate the additional
constraint on dressed spin polarization. The first and second order corrections to the order parameter are subsequently derived in
Sec.~\ref{subsec:pert_1st} and Sec.~\ref{subsec:pert_2nd}, respectively.

\subsection{General outline of the perturbative approach}
\label{subsec:pert_gen}
To perturbatively solve the time-independent Gross-Pitaevskii equation~\eqref{eq:soc_ti_gp_p}, we first perform a redefinition of the order parameter,
$\Psi_0(x') \to \me^{\mi k_c x'} \Psi_0(X)$, so that $\Psi_0$ is now a function of the single dimensionless variable $X = k_1 x'$ satisfying the
antiperiodicity condition $\Psi_0(X + \pi) = - \Psi_0(X)$ ($\Psi_0$ is thus $2\pi$-periodic in $X$). With these variable changes
Eq.~\eqref{eq:soc_ti_gp_p} takes the form
\begin{align}
&{} \left[h_X + \frac{\hbar\Omega_R}{2} \sigma_x
+ g_{dd} \left(\Psi_0^\dagger \Psi_0\right) + g_{ss} \left(\Psi_0^\dagger \sigma_z \Psi_0\right) \sigma_z\right] \Psi_0 \notag \\
&{} = (\mu_d + \mu_s p_X) \Psi_0 \, ,
\label{eq:pert_ti_gp_p}
\end{align}
where we have defined the operators $p_X = - \mi \nabla_X$ and
\begin{equation}
h_X = E_R \left(\alpha_1 p_X - \sigma_z\right)^2 + 2 E_R \alpha_c \left(\alpha_1 p_X - \sigma_z\right) + E_R \alpha_c^2 \, ,
\label{eq:pert_soc_ham}
\end{equation}
as well as the dimensionless wave vectors $\alpha_c = k_c / k_R$ and $\alpha_1 = k_1 / k_R$. At $\Omega_R = 0$, the uniform solutions of
Eq.~\eqref{eq:pert_ti_gp_p} take the form~\eqref{eq:zero_Psi}, only with $k_R x'$ replaced by $X$. For arbitrary $\alpha_c$ and $\alpha_1$, these
solutions have average energy density equal to
\begin{equation}
\begin{split}
\bar{\varepsilon}^{(0)} = {}&{} \bar{n} \bigg[E_R (\alpha_1-1)^2 + 2 E_R \alpha_c (\alpha_1-1) \px \\
&{} + E_R \alpha_c^2 + \frac{G_{dd}}{2} + \frac{G_{ss}}{2} \px^2 \bigg] \, ,
\end{split}
\label{eq:pert_zero_en}
\end{equation}
and from the thermodynamic relations~\eqref{eqs:soc_en_dens_der} one deduces
\begin{subequations}
\label{eqs:pert_zero_mu}
\begin{align}
\mu_d^{(0)} = {}&{} E_R (\alpha_1-1)^2 + E_R \alpha_c^2 + G_{dd} \, ,
\label{eq:pert_zero_mud} \\
\mu_s^{(0)} = {}&{} 2 E_R \alpha_c (\alpha_1-1) + G_{ss} \px \, .
\label{eq:pert_zero_mus}
\end{align}
\end{subequations}
Note that $\bar{\varepsilon}^{(0)}$ is minimized by the choice $\alpha_c = 0$ and $\alpha_1 = 1$, which leads to the results of Sec.~\ref{subsec:zero_order}.

Let us now insert the expansions~\eqref{eq:pert_exp_Psi}, \eqref{eq:pert_exp_mu_n}, and~\eqref{eq:pert_exp_mu_s} into Eq.~\eqref{eq:pert_ti_gp_p} and collect
terms of equal order in the small parameter $\hbar\Omega_R / 4 E_R$. At perturbative order $l \geq 1$, we obtain the recurrence relation
\begin{equation}
\begin{split}
&{} \Big[ E_R \left(\alpha_1 p_X - \sigma_z\right)^2 + \left(2 E_R \alpha_c - G_{ss} \px\right) \left(p_X - \sigma_z\right) \\
&{} - E_R \left(\alpha_1 - 1\right)^2 + \mathcal{L}_D \Big] \Psi_0^{(l)} + \mathcal{L}_C \Psi_0^{(l)*} \\
&{} = \left[ \mu_d^{(l)} + \mu_s^{(l)} p_X \right] \Psi_0^{(0)} - \mathcal{J}^{(l)} \, .
\end{split}
\label{eq:pert_gp}
\end{equation}
The matrices $\mathcal{L}_\mathrm{D}$ and $\mathcal{L}_\mathrm{C}$ are defined as
\begin{widetext}
\begin{subequations}
\label{eq:pert_LDC}
\begin{align}
\mathcal{L}_\mathrm{D}
&{} = \frac{1}{2}
\begin{pmatrix}
(1+\px) (G_{dd} + G_{ss}) & \sqrt{1 - \px^2} (G_{dd} - G_{ss}) \me^{2 \mi X} \\
\sqrt{1 - \px^2} (G_{dd} - G_{ss}) \me^{- 2 \mi X} & (1-\px) (G_{dd} + G_{ss}) 
\end{pmatrix}
\, ,
\label{eq:pert_LD} \\
\mathcal{L}_\mathrm{C}
&{} = \frac{1}{2}
\begin{pmatrix}
(1+\px) (G_{dd} + G_{ss}) \me^{2 \mi X} & \sqrt{1 - \px^2} (G_{dd} - G_{ss}) \\
\sqrt{1 - \px^2} (G_{dd} - G_{ss}) & (1-\px) (G_{dd} + G_{ss}) \me^{- 2 \mi X}
\end{pmatrix}
\, ,
\label{eq:pert_LC}
\end{align}
\end{subequations}
\end{widetext}
and the source term $\mathcal{J}^{(l)}$ reads
\begin{align}
\mathcal{J}^{(l)} = {}&{} \frac{\hbar\Omega_R}{2} \, \sigma_x \Psi_0^{(l-1)}
- \sum_{l_1=1}^{l-1} \left[ \mu_d^{(l_1)} + \mu_s^{(l_1)} p_X\right] \Psi_0^{(l-l_1)} \notag \\
&{} + g_{dd} \sum_{l_1,l_2,l_3=0}^{l-1}
\left[\Psi_0^{(l_1)\dagger}\Psi_0^{(l_2)}\right] \Psi_0^{(l_3)}
\delta_{l,l_1 + l_2 + l_3} \notag \\
&{} + g_{ss} \sum_{l_1,l_2,l_3=0}^{l-1}
\left[\Psi_0^{(l_1)\dagger} \sigma_z \Psi_0^{(l_2)}\right]
\sigma_z \Psi_0^{(l_3)} \delta_{l,l_1 + l_2 + l_3} \, .
\label{eq:pert_source}
\end{align}
In deriving Eqs.~\eqref{eq:pert_LDC}, we have set $\theta = \Delta\theta = 0$ in the expression~\eqref{eq:zero_Psi} for the unperturbed order parameter,
without loss of generality. Equation~\eqref{eq:pert_gp} provides a recursive method for determining the $l$th order correction to the order parameter,
$\Psi_0^{(l)}$, once all lower-order corrections are known. The latter are required to evaluate $\mathcal{J}^{(l)}$ and the correction to the average
energy density, $\bar{\varepsilon}^{(l)}$, from which one obtains $\mu_d^{(l)}$ and $\mu_s^{(l)}$ using again the relations~\eqref{eqs:soc_en_dens_der}.
Specifically, $\bar{\varepsilon}^{(l)}$ can be computed inserting the expansion~\eqref{eq:pert_exp_Psi} into the energy~\eqref{eq:soc_en} divided by $V$.
After singling out the term of order $l$, changing integration variable from $x$ to $X$, and carrying out the trivial integration over $y$ and $z$, one
gets an expressions that depends on the yet unknown $\Psi_0^{(l)}$. However, this dependence appears only through two integrals, which can be expressed
using lower order quantities. These are
\begin{equation}
\begin{split}
&{} \int_{-\pi}^{\pi} \dif X \left[ \Psi_0^{(0)\dagger} \Psi_0^{(l)} + \Psi_0^{(l)\dagger} \Psi_0^{(0)} \right] \\
&{} = - \sum_{l_1 = 1}^{l-1} \int_{-\pi}^{\pi} \dif X \, \Psi_0^{(l_1)\dagger} \Psi_0^{(l-l_1)}
\end{split}
\label{eq:pert_cons_N}
\end{equation}
and
\begin{equation}
\begin{split}
&{} \int_{-\pi}^{\pi} \dif X
\left[ \Psi_0^{(0)\dagger} p_X \Psi_0^{(l)} + \Psi_0^{(l)\dagger} p_X \Psi_0^{(0)} \right] \\
&{} = - \sum_{l_1 = 1}^{l-1} \int_{-\pi}^{\pi} \dif X \, \Psi_0^{(l_1)\dagger} p_X \Psi_0^{(l-l_1)} \, .
\end{split}
\label{eq:pert_cons_p}
\end{equation}
For simplicity, here we are restricting integration over a single period of the order parameter, from $X = - \pi$ to $X = \pi$.
Equations~\eqref{eq:pert_cons_N} and~\eqref{eq:pert_cons_p} are obtained by imposing the conservation of particle number [Eq.~\eqref{eq:soc_part_numb}]
and momentum [Eq.~\eqref{eq:soc_can_mom}] at each perturbative order. The relation $p_X \Psi_0^{(0)} = \sigma_z \Psi_0^{(0)}$ must also be employed in the
calculation of $\bar{\varepsilon}^{(l)}$.

Aside from the evaluation of $\mu_d^{(l)}$ and $\mu_s^{(l)}$, the knowledge of all the energy density corrections up to order $l$ is also crucial to
determine the optimal values of the dimensionless wave vectors $\alpha_c$ and $\alpha_1$ up to the same order. This is achieved by imposing the two
stationarity conditions $\frac{\partial}{\partial \alpha_c} \sum_{l_1=0}^l \bar{\varepsilon}^{(l_1)} = 0$ and $\frac{\partial}{\partial \alpha_1}
\sum_{l_1=0}^l \bar{\varepsilon}^{(l_1)} = 0$. We then seek solutions in the form of perturbative expansions, $\alpha_c = \sum_{l_1=0}^l
\alpha_c^{(l_1)}$ and $\alpha_1 = \sum_{l_1=0}^l \alpha_1^{(l_1)}$, where the unperturbed values $\alpha_c^{(0)} = 0$ and $\alpha_1^{(0)} = 1$
coincide with those obtained previously. This self-consistent procedure allows the corrections $\alpha_c^{(l)}$ and $\alpha_1^{(l)}$ to be
determined iteratively, order by order.

The general solution of the nonhomogeneous linear equation~\eqref{eq:pert_gp} can be constructed by adding a particular solution to the general solution
$\Psi_\mathrm{Hom}^{(l)}$ of the associated homogeneous equation. The latter is a linear combination of eight linearly independent functions with arbitrary
coefficients. As in the $\px = 0$ case discussed in Ref.~\cite{Martone2021b}, six of these functions violate the periodicity condition of the order
parameter and must therefore be excluded. This leaves only two admissible functions, and $\Psi_\mathrm{Hom}^{(l)}$ reduces to the form
\begin{equation}
\Psi_\mathrm{Hom}^{(l)} = \mi \left[ \theta^{(l)} + \Delta\theta^{(l)} p_X \right] \Psi_0^{(0)} \, ,
\label{eq:pert_hom_gen_sol}
\end{equation}
where $\theta^{(l)}$ and $\Delta\theta^{(l)}$ are undetermined constants. To maintain consistency with the perturbative expansion, both coefficients must
scale as $(\hbar \Omega_R / 4 E_R)^l$. The two terms in Eq.~\eqref{eq:pert_hom_gen_sol} correspond to infinitesimal shifts in the condensate phase
and in the position of density fringes, respectively, which reflects the spontaneous breaking of global $\mathrm{U}(1)$ and translation symmetry in the
stripe phase. To uniquely determine $\theta^{(l)}$ and $\Delta\theta^{(l)}$, we impose the following orthogonality conditions on the order-$l$ correction
to the order parameter:
\begin{subequations}
\label{eq:pert_Psi_ph_off}
\begin{align}
&{} \int_{-\pi}^{\pi} \dif X \, \Im\left[\Psi_0^{(0)\dagger} \Psi_0^{(l)}\right] = 0 \, ,
\label{eq:pert_Psi_ph} \\
&{} \int_{-\pi}^{\pi} \dif X \, \Im\left[\Psi_0^{(0)\dagger} p_X \Psi_0^{(l)}\right] = 0 \, .
\label{eq:pert_Psi_off}
\end{align}
\end{subequations}

The function $\Psi_0^{(l)}$ computed following the above prescriptions is a sum of plane-wave components whose wave vectors (in units of $k_1$) are all odd
integers from $-(2 l + 1)$ to $2 l + 1$. Thus, the number of harmonics increases with the perturbative order, in agreement with the Bloch-wave structure of
the exact order parameter, Eq.~\eqref{eq:soc_stripe_wf_bl}. In the next sections, we explicitly compute the corrections at first and second order in
$\hbar \Omega_R / 4 E_R$.

\subsection{First order results}
\label{subsec:pert_1st}
For $l = 1$, the source term~\eqref{eq:pert_source} receives contributions solely from the Raman coupling, yielding
\begin{equation}
\mathcal{J}^{(1)}
= 
\begin{pmatrix}
0 \\
\mathcal{J}_{+1,\downarrow}^{(1)}(\px)
\end{pmatrix}
\me^{\mi X}
+
\begin{pmatrix}
\mathcal{J}_{-1,\uparrow}^{(1)}(\px) \\
0
\end{pmatrix}
\me^{- \mi X} \, ,
\label{eq:pert_1st_source}
\end{equation}
with
\begin{equation}
\mathcal{J}_{+1,\downarrow}^{(1)}(\px) = \mathcal{J}_{-1,\uparrow}^{(1)}(-\px) = \sqrt{\frac{\bar{n}(1+\px)}{2}} \frac{\hbar \Omega_R}{2} \, .
\label{eq:pert_1st_source_coeff}
\end{equation}
In addition, following the procedure outlined in Sec.~\ref{subsec:pert_gen}, one immediately finds that $\bar{\varepsilon}^{(1)} = 0$, meaning that $\mu_d^{(1)}
= \mu_s^{(1)} = 0$ and $\alpha_c^{(1)} = \alpha_1^{(1)} = 0$. The first order correction to the order parameter, satisfying both Eq.~\eqref{eq:pert_gp} and the
orthogonality conditions~\eqref{eq:pert_Psi_ph_off}, takes the form
\begin{equation}
\begin{split}
\Psi_0^{(1)} = {}&{}
\begin{pmatrix}
\tilde{\Psi}_{+3,\uparrow}^{(1)}(\px) \\
0
\end{pmatrix}
\me^{3 \mi X}
+
\begin{pmatrix}
0 \\
\tilde{\Psi}_{+1,\downarrow}^{(1)}(\px)
\end{pmatrix}
\me^{\mi X}
\\
{}&{}
+
\begin{pmatrix}
\tilde{\Psi}_{-1,\uparrow}^{(1)}(\px) \\
0
\end{pmatrix}
\me^{- \mi X}
+
\begin{pmatrix}
0 \\
\tilde{\Psi}_{-3,\downarrow}^{(1)}(\px)
\end{pmatrix}
\me^{- 3 \mi X} \, .
\end{split}
\label{eq:pert_1st_Psi_sol}
\end{equation}
To determine the coefficients in the above expression, we insert the Ansatz~\eqref{eq:pert_1st_Psi_sol} into Eq.~\eqref{eq:pert_gp}, group terms with the same
oscillatory behavior, and equate coefficients on both sides of the equation. The final results are functions of $\px$, $\alpha_c$, and $\alpha_1$. Here we provide
the explicit expressions when $\alpha_c$ and $\alpha_1$ are taken equal to the optimal values given above:
\begin{subequations}
\label{eq:pert_1st_Psi_coeff}
\begin{align}
\tilde{\Psi}_{+1,\downarrow}^{(1)}(\px) &{} = \tilde{\Psi}_{-1,\uparrow}^{(1)}(-\px)
= \sqrt{\frac{\bar{n}(1+\px)}{2}} \frac{N_1^{(1)}}{2 D_1} \frac{\hbar \Omega_R}{4 E_R} \, ,
\label{eq:pert_1st_Psi_coeff_1} \\
\begin{split}
\tilde{\Psi}_{+3,\uparrow}^{(1)}(\px) &{} = \tilde{\Psi}_{-3,\downarrow}^{(1)}(-\px) \\
&{} = \sqrt{\frac{\bar{n}(1-\px)}{2}} \frac{(1+\px) N_3^{(1)}}{2 D_1} \frac{\hbar \Omega_R}{4 E_R} \, .
\end{split}
\label{eq:pert_1st_Psi_coeff_3}
\end{align}
\end{subequations}
The explicit forms of $D_1$, $D_2$, $N_1^{(1)}$, and $N_3^{(1)}$ are given in Appendix~\ref{sec:pert_formulas}. Once $\Psi_0^{(1)}$ is known, any observable can be
computed to first order in $\hbar \Omega_R / 4 E_R$. Among those discussed in Sec.~\ref{subsec:res_obs}, only the density profiles [see Eqs.~\eqref{eq:res_tot_dens},
\eqref{eq:res_spin_dens}, \eqref{eq:res_dens_coeff_1}, and~\eqref{eq:res_spin_coeff_1}] and the associated contrasts exhibit nonzero first order corrections.

\subsection{Second order results}
\label{subsec:pert_2nd}
At second order in the Raman coupling ($l=2$), the source term takes the form
\begin{equation}
\begin{split}
\mathcal{J}^{(2)} = {}&{}
\begin{pmatrix}
\mathcal{J}_{+5,\uparrow}^{(2)}(\px) \\
0
\end{pmatrix}
\me^{5 \mi X}
+
\begin{pmatrix}
0 \\
\mathcal{J}_{+3,\downarrow}^{(2)}(\px)
\end{pmatrix}
\me^{3 \mi X}
\\
{}&{}
+
\begin{pmatrix}
\mathcal{J}_{+1,\uparrow}^{(2)}(\px) \\
0
\end{pmatrix}
\me^{\mi X}
+
\begin{pmatrix}
0 \\
\mathcal{J}_{-1,\downarrow}^{(2)}(\px)
\end{pmatrix}
\me^{- \mi X}
\\
{}&{}
+
\begin{pmatrix}
\mathcal{J}_{-3,\uparrow}^{(2)}(\px) \\
0
\end{pmatrix}
\me^{- 3 \mi X}
+
\begin{pmatrix}
0 \\
\mathcal{J}_{-5,\downarrow}^{(2)}(\px)
\end{pmatrix}
\me^{- 5 \mi X} \, .
\end{split}
\label{eq:pert_2nd_source}
\end{equation}
The coefficients in this expression can be derived from the general formula~\eqref{eq:pert_source}, using the results for the first order corrections to the various
quantities. Applying the method of Sec.~\ref{subsec:pert_gen}, one obtains the second order corrections to the average energy density $\bar{\varepsilon}^{(2)}$,
the mean chemical potentials $\mu_d^{(2)}$, the chemical potential semi-difference $\mu_s^{(2)}$, and the dimensionless wave vectors $\alpha_c^{(2)}$ and $\alpha_1^{(2)}$.
These are all reported in the main text [see second term on the right hand side of Eqs.~\eqref{eq:res_en}, \eqref{eq:res_mu_d}, \eqref{eq:res_mu_s}, \eqref{eq:res_kc},
and~\eqref{eq:res_k1}]. The second order correction to the order parameter, derived in the same manner as the first order one (see Sec.~\ref{subsec:pert_1st}), reads
\begin{equation}
\begin{split}
\Psi_0^{(2)} = {}&{}
\begin{pmatrix}
\tilde{\Psi}_{+5,\uparrow}^{(2)}(\px) \\
0
\end{pmatrix}
\me^{5 \mi X}
+
\begin{pmatrix}
0 \\
\tilde{\Psi}_{+3,\downarrow}^{(2)}(\px)
\end{pmatrix}
\me^{3 \mi X}
\\
{}&{}
+
\begin{pmatrix}
\tilde{\Psi}_{+1,\uparrow}^{(2)}(\px) \\
0
\end{pmatrix}
\me^{\mi X}
+
\begin{pmatrix}
0 \\
\tilde{\Psi}_{-1,\downarrow}^{(2)}(\px)
\end{pmatrix}
\me^{- \mi X}
\\
{}&{}
+
\begin{pmatrix}
\tilde{\Psi}_{-3,\uparrow}^{(2)}(\px) \\
0
\end{pmatrix}
\me^{- 3 \mi X}
+
\begin{pmatrix}
0 \\
\tilde{\Psi}_{-5,\downarrow}^{(2)}(\px)
\end{pmatrix}
\me^{- 5 \mi X} \, .
\end{split}
\label{eq:pert_2nd_Psi_sol}
\end{equation}
As for the first order corrections, we give the explicit expressions of the coefficients in Eq.~\eqref{eq:pert_2nd_Psi_sol} only for $\alpha_c$ and $\alpha_1$
equal to their optimal values:
\begin{subequations}
\label{eq:pert_2nd_Psi_coeff}
\begin{align}
\begin{split}
\tilde{\Psi}_{+1,\uparrow}^{(2)}(\px) &{} = \tilde{\Psi}_{-1,\downarrow}^{(2)}(-\px) \\
&{} = \sqrt{\frac{\bar{n}(1+\px)}{2}} \frac{N_1^{(2)}}{4 D_1^2}
\left(\frac{\hbar\Omega_R}{4 E_R}\right)^2 \, ,
\end{split}
\label{eq:pert_2nd_Psi_coeff_1} \\
\begin{split}
\tilde{\Psi}_{+3,\downarrow}^{(2)}(\px) &{} = \tilde{\Psi}_{-3,\uparrow}^{(2)}(-\px) \\
&{} = - \sqrt{\frac{\bar{n}(1-\px)}{2}} \frac{(1+\px) N_3^{(2)}}{4 D_1^2 D_2}
\left( \frac{\hbar \Omega_R}{4 E_R} \right)^2 \, ,
\end{split}
\label{eq:pert_2nd_Psi_coeff_3} \\
\begin{split}
\tilde{\Psi}_{+5,\uparrow}^{(2)}(\px) &{} = \tilde{\Psi}_{-5,\downarrow}^{(2)}(-\px) \\
&{} = \sqrt{\frac{\bar{n}(1+\px)}{2}} \frac{(1-\px^2) N_5^{(2)}}{4 D_1^2 D_2}
\left( \frac{\hbar \Omega_R}{4 E_R} \right)^2 \, .
\end{split}
\label{eq:pert_2nd_Psi_coeff_5}
\end{align}
\end{subequations}
The quantities $D_1$, $D_2$, $N_1^{(2)}$, $N_3^{(2)}$, and $N_5^{(2)}$ are given in Appendix~\ref{sec:pert_formulas}. The structure of
Eqs.~\eqref{eq:pert_2nd_Psi_coeff} respects the orthogonality constraints~\eqref{eq:pert_Psi_ph_off}. With the expression for $\Psi_0^{(2)}$
at hand, the second order corrections to observables can be readily computed. In addition to those mentioned earlier, in
Sec.~\ref{subsec:res_obs} of the main text we present the results for the total and bare spin density profiles [see Eqs.~\eqref{eq:res_tot_dens},
\eqref{eq:res_spin_dens}, \eqref{eq:res_dens_coeff_2}, and~\eqref{eq:res_spin_coeff_2}], the average bare spin polarization density $\bar{s}_z$
[Eq.~\eqref{eq:res_polar}], and the fringe velocity $v$ [Eq.~\eqref{eq:res_vel}]. Finally, it is worth noting that all the perturbative expressions
reported in this appendix reduce to those of Ref.~\cite{Martone2021b} when $\px = 0$.

\section{Coefficients of perturbative formulas}
\label{sec:pert_formulas}
In this Appendix we give the explicit forms of the coefficients introduced in Secs.~\ref{subsec:res_obs}, \ref{subsec:pert_1st},
and~\ref{subsec:pert_2nd}. These coefficients appear in various perturbative formulas.

The two quantities appearing in the denominators of all the perturbative expressions are
\begin{align}
D_1 = {}&{} 4 E_R^2 \left(2 E_R+G_{ss}\right) \left(2 E_R+G_{dd}\right) - 2 E_R G_{ss} \notag \\
&{} \times \left[\left(4 E_R+G_{ss}\right) G_{ss} + \left(2 E_R+G_{ss}\right) G_{dd}\right] \px^2 \notag \\
&{} + G_{ss}^4 \px^4
\label{eq:pert_form_D1}
\end{align}
and
\begin{align}
D_2 = {}&{} 4 E_R^2 \left(8 E_R+G_{ss}\right) \left(8 E_R+G_{dd}\right) - 2 E_R G_{ss} \notag \\
&{} \times \left[\left(16 E_R+G_{ss}\right) G_{ss} + \left(2 E_R+G_{ss}\right) G_{dd}\right] \px^2 \notag \\
&{} + G_{ss}^4 \px^4 \, .
\label{eq:pert_form_D2}
\end{align}
The coefficients entering the numerators of the observables computed in Sec.~\ref{subsec:res_obs} are
\begin{align}
N_{\tilde{n}_2} = {}&{} 48 E_R^5 (2 E_R+G_{ss})^2 (8 E_R+G_{ss}) (4 E_R+G_{dd}) G_{dd} \notag \\
&{} + 16 E_R^4 G_{ss} \big[ E_R \left(16 E_R^2 - 16 E_R G_{ss} - 3 G_{ss}^2\right) G_{ss}^2 \notag \\
&{} + \left(4 E_R^3 - 48 E_R^2 G_{ss} - 31 E_R G_{ss}^2 - 3 G_{ss}^3\right) G_{ss} G_{dd} \notag \\
&{} + \left(12 E_R^3 + 8 E_R^2 G_{ss} - E_R G_{ss}^2 - G_{ss}^3\right) G_{dd}^2 \big] \px^2 \notag \\
&{} - 4 E_R^3 G_{ss}^3 \Big[ \left(24 E_R^2 - 20 E_R G_{ss} - 3 G_{ss}^2\right) G_{ss}^2 \notag \\
&{} + 4 \left(6 E_R^2 + E_R G_{ss} - G_{ss}^2\right) G_{ss} G_{dd} \notag \\
&{} + \left(2 E_R+G_{ss}\right)^2 G_{dd}^2 \Big] \px^4 \notag \\
&{} + 4 E_R^2 G_{ss}^6 [ \left(3 E_R-G_{ss}\right) G_{ss} \notag \\
&{} + \left(2 E_R+G_{ss}\right) G_{dd} ] \px^6 - E_R G_{ss}^9 \px^8 \, ,
\label{eq:pert_form_N_n2}
\end{align}
\begin{align}
N_{\tilde{s}_{z,2}} = {}&{} 16 E_R^4 (2 E_R+G_{ss}) \big[ 128 E_R^3 G_{ss}^2 \notag \\
&{} + 32 E_R^2 \left(7 E_R + 3 G_{ss}\right) G_{ss} G_{dd} \notag \\
&{} + 2 E_R \left(72 E_R^2 + 58 E_R G_{ss} + 9 G_{ss}^2\right) G_{dd}^2 \notag \\
&{} + \left(12 E_R+G_{ss}\right) \left(2 E_R+G_{ss}\right) G_{dd}^3 \big] \notag \\
&{} - 8 E_R^3 G_{ss}^2 \Big[ 32 E_R^2 \left(7 E_R + 3 G_{ss}\right) G_{ss}^2 \notag \\
&{} + 4 E_R \left(62 E_R^2 + 53 E_R G_{ss} + 9 G_{ss}^2\right) G_{ss} G_{dd} \notag \\
&{} + \left(56 E_R^3 + 108 E_R^2 G_{ss} + 46 E_R G_{ss}^2 + 3 G_{ss}^3\right) G_{dd}^2 \notag \\
&{} + \left(2 E_R+G_{ss}\right)^2 G_{dd}^3 \Big] \px^2 \notag \\
&{} + 4 E_R^2 G_{ss}^4 \big[ 18 E_R \left(4 E_R+G_{ss}\right) G_{ss}^2 \notag \\
&{} + \left(56 E_R^2 + 44 E_R G_{ss} + 3 G_{ss}^2\right) G_{ss} G_{dd} \notag \\
&{} + \left(4 E_R^2 + 8 E_R G_{ss} + 3 G_{ss}^2\right) G_{dd}^2 \big] \px^4 \notag \\
&{} - 2 E_R G_{ss}^7 [ \left(14 E_R + G_{ss}\right) G_{ss} \notag \\
&{} + \left(4 E_R + 3 G_{ss}\right) G_{dd} ] \px^6 + G_{ss}^{10} \px^8 \, ,
\label{eq:pert_form_N_s2}
\end{align}
\begin{align}
N_{k_c} = {}&{}
4 E_R^2 (2 E_R+G_{ss}) \big[\left(4 E_R+G_{ss}\right) G_{ss} \notag \\
&{} + 2 \left(E_R+G_{ss}\right) G_{dd} \big] \notag \\
&{} - 2 E_R G_{ss}^2 \big[\left(4 E_R+3 G_{ss}\right) G_{ss} \notag \\
&{} - \left(2 E_R+G_{ss}\right) G_{dd}\big] \px^2 - G_{ss}^5 \px^4 \, ,
\label{eq:pert_form_N_kc}
\end{align}
\begin{align}
N_{k_1} = {}&{}
4 E_R^2 (2 E_R+G_{ss})^2 \left(4 E_R^2+2 E_R G_{dd} + G_{dd}^2\right) \notag \\
&{} - 4 E_R G_{ss} \big[ 2 E_R^2 G_{ss}^2 - \big(4 E_R^3 - 3 E_R G_{ss}^2 \notag \\
&{} - G_{ss}^3\big) G_{dd} \big] \px^2 - G_{ss}^4 \left(4 E_R^2 - 2 E_R G_{ss} - G_{ss}^2\right) \px^4 \, ,
\label{eq:pert_form_N_k1}
\end{align}
\begin{align}
N_{\bar{s}_z} = {}&{} 4 E_R^2 \big[2 E_R \left(4 E_R^2+6 G_{ss} E_R+G_{ss}^2\right) \notag \\
&{} + 2 E_R \left(4 E_R+3 G_{ss}\right) G_{dd} + \left(2 E_R+G_{ss}\right) G_{dd}^2\big] \notag \\
&{} - 4 E_R G_{ss} \left[3 E_R G_{ss}^2 - \left(2 E_R^2-G_{ss}^2\right) G_{dd}\right] \px^2 \notag \\
&{} - (2 E_R-G_{ss}) G_{ss}^4 \px^4 \, ,
\label{eq:pert_form_N_sz}
\end{align}
\begin{align}
N_{\bar{\varepsilon}} = {}&{} 2 E_R \left(2 E_R+G_{ss}\right) \left(4 E_R+G_{dd}\right) \notag \\
&{} + \left[\left(8 E_R^2-G_{ss}^2\right) G_{ss} + 2 E_R \left(2 E_R+G_{ss}\right) G_{dd}\right] \px^2 \notag \\
&{} - G_{ss}^3 \px^4 \, ,
\label{eq:pert_form_N_eps}
\end{align}
\begin{align}
N_{\mu_d} = {}&{} - 4 E_R^3 \left(2 E_R+G_{ss}\right)^2 \left(8 E_R^2+4 E_R G_{dd}+G_{dd}^2\right) \notag \\
&{} + 4 E_R^2 \big[ E_R \left(24 E_R^2+20 E_R G_{ss}+3 G_{ss}^2\right) G_{ss}^2 \notag \\
&{} + \left(16 E_R^3+24 E_R^2 G_{ss}+10 E_R G_{ss}^2+G_{ss}^3\right) G_{ss} G_{dd} \notag \\
&{} + \left(4 E_R^3+8 E_R^2 G_{ss}+5 E_R G_{ss}^2+G_{ss}^3\right) G_{dd}^2 \big] \px^2 \notag \\
&{} - E_R G_{ss}^3 \big[\left(20 E_R^2+16 E_R G_{ss}+G_{ss}^2\right) G_{ss} \notag \\
&{} + \left(8 E_R^2+12 E_R G_{ss}+4 G_{ss}^2\right) G_{dd}\big] \px^4 \notag \\
&{} + \left(E_R+G_{ss}\right) G_{ss}^6 \px^6 \, ,
\label{eq:pert_form_N_mu_d}
\end{align}
\begin{align}
N_{\mu_s} = {}&{} 4 E_R^2 \Big[2 E_R \left(8 E_R^2+8 E_R G_{ss}+G_{ss}^2\right) G_{ss} \notag \\
&{} + 4 E_R \left(2 E_R^2 +5 E_R G_{ss}+2 G_{ss}^2\right) G_{dd} \notag \\
&{} + \left(2 E_R+G_{ss}\right)^2 G_{dd}^2\Big] \notag \\
&{} - 4 E_R G_{ss}^3 [4 E_R \left(E_R+G_{ss}\right) \notag \\
&{} + \left(2 E_R+G_{ss}\right) G_{dd}] \px^2 + G_{ss}^6 \px^4 \, ,
\label{eq:pert_form_N_mu_s}
\end{align}
and
\begin{align}
N_v = {}&{} 2 E_R G_{ss} \left(4 E_R^2+6 E_R G_{ss}+G_{ss}^2\right) (2 E_R+G_{ss}) \notag \\
&{} + 2 E_R (2 E_R + 3 G_{ss}) (2 E_R + G_{ss})^2 G_{dd} \notag \\
&{} + 2 E_R (2 E_R + G_{ss})^2 G_{dd}^2 \notag \\
&{} - \Big[2 G_{ss}^3 \left(4 E_R^2+5 E_R G_{ss}+2 G_{ss}^2\right) \notag \\
&{} + (2 E_R+G_{ss})^2 G_{ss}^2 G_{dd} \Big] \px^2 + G_{ss}^5 \px^4 \, .
\label{eq:pert_form_N_v}
\end{align}
In the numerators of the first order corrections to the Fourier expansion coefficients of the order parameter, evaluated in Appendix~\ref{subsec:pert_1st},
one has
\begin{align}
N_1^{(1)} = {}&{} - 2 E_R^2 \left(4 E_R+G_{dd}\right) \left(2 E_R+G_{ss}\right) - 2 E_R^2 \notag \\
&{} \times \left[\left(4 E_R+G_{ss}\right) G_{ss} + \left(2 E_R+G_{ss}\right) G_{dd}\right] \px \notag \\
&{} + E_R \left(2 E_R+G_{ss}\right) G_{ss}^2 \px^2 + E_R G_{ss}^3 \px^3
\label{eq:pert_form_N_1_1}
\end{align}
and
\begin{align}
N_3^{(1)} = {}&{} 2 E_R^2 \left(2 E_R+G_{ss}\right) G_{dd} \notag \\
&{} - 2 E_R^2 G_{ss}^2 \px - E_R G_{ss}^3 \px^2 \, .
\label{eq:pert_form_N_3_1}
\end{align}
Finally, the second order corrections to the Fourier expansion coefficients of the order parameter (see Appendix~\ref{subsec:pert_2nd}) depend on
the following quantities:
\begin{align}
N_1^{(2)} = {}&{} - 4 E_R^4 (2 E_R + G_{ss})^2 \left(8 E_R^2 + 4 E_R G_{dd} + G_{dd}^2\right) \notag \\
&{} - 8 E_R^4 (2 E_R + G_{ss}) \big[2 E_R \left(4 E_R + G_{ss}\right) G_{ss} \notag \\
&{} + \left(4 E_R^2 + 4 E_R G_{ss} - G_{ss}^2\right) G_{dd} \notag \\
&{} + \left(2 E_R + G_{ss}\right) G_{dd}^2 \big] \px \notag \\
&{} + 4 E_R^3 \Big[E_R \left(4 E_R + G_{ss}\right) G_{ss}^3 - \big(8 E_R^3 + 8 E_R^2 G_{ss} \notag \\
&{} - G_{ss}^3\big) G_{ss} G_{dd} + E_R \left(2 E_R + G_{ss}\right)^2 G_{dd}^2 \Big] \px^2 \notag \\
&{} + 4 E_R^3 G_{ss}^3 \big[8 E_R^2 + 4 E_R G_{ss} - G_{ss}^2 \notag \\
&{} + 2 \left(2 E_R + G_{ss}\right) G_{dd}\big] \px^3 \notag \\
&{} - E_R^2 G_{ss}^3 (2 E_R + G_{ss}) \big(G_{ss}^2 - 6 E_R G_{ss} \notag \\
&{} + 4 E_R G_{dd}\big) \px^4 - 2 E_R^2 G_{ss}^6 \px^5 + E_R^2 G_{ss}^6 \px^6 \, ,
\label{eq:pert_form_N_1_2}
\end{align}
\begin{align}
N_3^{(2)} = {}&{} 16 E_R^7 \left(2 E_R+G_{ss}\right)^2 \left(8 E_R+G_{ss}\right) (16 E_R \notag \\
&{} +5 G_{dd}) G_{dd} - 8 E_R^6 (2 E_R+G_{ss}) \big[ 32 E_R^2 G_{ss}^3 \notag \\
&{} - 8 E_R \left(48 E_R^2 + 15 E_R G_{ss} - G_{ss}^2\right) G_{ss} G_{dd} \notag \\
&{} - \left(160 E_R^3 + 168 E_R^2 G_{ss} + 38 E_R G_{ss}^2 + G_{ss}^3\right) G_{dd}^2 \notag \\
&{} - \left(16 E_R+G_{ss}\right) \left(2 E_R+G_{ss}\right) G_{dd}^3 \big] \px \notag \\
&{} + 8 E_R^6 G_{ss} \Big[ 4 E_R \left(16 E_R^2 - 4 E_R G_{ss} - 3 G_{ss}^2\right) G_{ss}^2 \notag \\
&{} + \big(48 E_R^3 - 92 E_R^2 G_{ss} - 74 E_R G_{ss}^2 \notag \\
&{} - 11 G_{ss}^3\big) G_{ss} G_{dd} + 2 E_R \big(36 E_R^2 + 28 E_R G_{ss} \notag \\
&{} + 5 G_{ss}^2\big) G_{dd}^2 + 2 \left(2 E_R+G_{ss}\right)^2 G_{dd}^3 \Big] \px^2 \notag \\
&{} + 4 E_R^5 G_{ss} \big[2 E_R \left(48 E_R^2 + 12 E_R G_{ss} + 7 G_{ss}^2\right) G_{ss}^3 \notag \\
&{} - 2 \big(44 E_R^3 + 80 E_R^2 G_{ss} + 26 E_R G_{ss}^2 \notag \\
&{} + G_{ss}^3\big) G_{ss}^2 G_{dd} - \big(8 E_R^3 + 76 E_R^2 G_{ss} + 42 E_R G_{ss}^2 \notag \\
&{} + 3 G_{ss}^3\big) G_{ss} G_{dd}^2 + \big(8 E_R^3 + 4 E_R^2 G_{ss} - 2 E_R G_{ss}^2 \notag \\
&{} - G_{ss}^3\big) G_{dd}^3 \big] \px^3 + 4 E_R^5 G_{ss}^2 \Big[ 2 \left(5 E_R + 3 G_{ss}\right) G_{ss}^4 \notag \\
&{} + \left(20 E_R^2 + 16 E_R G_{ss} + G_{ss}^2\right) G_{ss}^2 G_{dd} \notag \\
&{} + \left(4 E_R^2 - 4 E_R G_{ss} - 3 G_{ss}^2\right) G_{ss} G_{dd}^2 \notag \\
&{} + \left(2 E_R + G_{ss}\right)^2 G_{dd}^3 \Big] \px^4 \notag \\
&{} - 2 E_R^4 G_{ss}^4 \big[ \left(48 E_R^2 - 6 E_R G_{ss} - G_{ss}^2\right) G_{ss}^2 \notag \\
&{} - \left(4 E_R^2 + 22 E_R G_{ss} + 3 G_{ss}^2\right) G_{ss} G_{dd} \notag \\
&{} + 2 \left(4 E_R^2 - G_{ss}^2\right) G_{dd}^2 \big] \px^5 \notag \\
&{} - 2 E_R^4 G_{ss}^5 \big[(18 E_R+G_{ss}) G_{ss}^2 \notag \\
&{} + 4 E_R G_{ss} G_{dd} + 2 (2 E_R+G_{ss}) G_{dd}^2\big] \px^6 \notag \\
&{} - E_R^3 G_{ss}^7 \left[G_{ss}^2 - \left(2 E_R-G_{ss}\right) G_{dd}\right] \px^7 \notag \\
&{} + E_R^3 G_{ss}^8 (G_{ss} + G_{dd}) \px^8 \, ,
\label{eq:pert_form_N_3_2}
\end{align}
and
\begin{align}
N_5^{(2)} = {}&{} 16 E_R^6 (2 E_R+G_{ss})^2 (8 E_R+G_{ss}) (5 E_R+G_{dd}) G_{dd}^2 \notag \\
&{} - 8 E_R^6 G_{ss} (2 E_R+G_{ss}) \big[ 32 E_R^2 G_{ss}^2 \notag \\
&{} + 8 E_R \left(15 E_R + 4 G_{ss}\right) G_{ss} G_{dd} \notag \\
&{} + \left(28 E_R^2 + 32 E_R G_{ss} + 5 G_{ss}^2\right) G_{dd}^2 \notag \\
&{} + \left(2 E_R+G_{ss}\right) G_{dd}^3 \big] \px \notag \\
&{} + 8 E_R^5 G_{ss} \big[ 4 E_R^2 \left(4 E_R-G_{ss}\right) G_{ss}^3 \notag \\
&{} - E_R \left(52 E_R^2 + 50 E_R G_{ss} + 9 G_{ss}^2\right) G_{ss}^2 G_{dd} \notag \\
&{} - \big(28 E_R^3 + 64 E_R^2 G_{ss} + 31 E_R G_{ss}^2 \notag \\
&{} + 3 G_{ss}^3\big) G_{ss} G_{dd}^2 \notag \\
&{} + \big(4 E_R^3 - 3 E_R G_{ss}^2 - G_{ss}^3\big) G_{dd}^3 \big] \px^2 \notag \\
&{} + 4 E_R^5 G_{ss}^2 \Big[ 2 E_R \left(48 E_R + 13 G_{ss}\right) G_{ss}^3 \notag \\
&{} + 2 \left(66 E_R^2 + 45 E_R G_{ss} + 5 G_{ss}^2\right) G_{ss}^2 G_{dd} \notag \\
&{} + \left(20 E_R^2 + 24 E_R G_{ss} + 7 G_{ss}^2\right) G_{ss} G_{dd}^2 \notag \\
&{} + \left(2 E_R+G_{ss}\right)^2 G_{dd}^3 \Big] \px^3 + 4 E_R^4 G_{ss}^5 \big[ 4 E_R G_{ss}^2 \notag \\
&{} + \left(26 E_R^2 + 25 E_R G_{ss} + 3 G_{ss}^2\right) G_{dd} \notag \\
&{} + 3 \left(2 E_R+G_{ss}\right) G_{dd}^2 \big] \px^4 \notag \\
&{} - 2 E_R^4 G_{ss}^5 \big[ \left(42 E_R + 5 G_{ss}\right) G_{ss}^2 \notag \\
&{} + \left(20 E_R + 11 G_{ss}\right) G_{ss} G_{dd} \notag \\
&{} + 2 \left(2 E_R+G_{ss}\right) G_{dd}^2 \big] \px^5 \notag \\
&{} - 2 E_R^3 G_{ss}^7 [ \left(6 E_R+G_{ss}\right) G_{ss} \notag \\
&{} + 3 \left(E_R+G_{ss}\right) G_{dd} ] \px^6 \notag \\
&{} + E_R^3 G_{ss}^8 (5 G_{ss} + G_{dd}) \px^7 + E_R^2 G_{ss}^{10} \px^8 \, .
\label{eq:pert_form_N_5_2}
\end{align}


\begin{thebibliography}{99}

\bibitem{Balibar_review}
S.~Balibar,
The enigma of supersolidity,
\href{https://doi.org/10.1038/nature08913}
{Nature (London) \textbf{464}, 176 (2010)}.

\bibitem{Boninsegni_review}
M.~Boninsegni and N.~V. Prokof'ev,
\textit{Colloquium:} Supersolids: What and where are they?,
\href{https://doi.org/10.1103/RevModPhys.84.759}
{Rev. Mod. Phys. \textbf{84}, 759 (2012)}.

\bibitem{Boettcher_review}
F.~B\"{o}ttcher, J.-N. Schmidt, J.~Hertkorn, K.~S.~H. Ng,
S.~D. Graham, M.~Guo, T.~Langen, and T.~Pfau,
New states of matter with fine-tuned interactions:
quantum droplets and dipolar supersolids,
\href{https://doi.org/10.1088/1361-6633/abc9ab}
{Rep. Prog. Phys. \textbf{84}, 012403 (2021)}.

\bibitem{Recati_review}
A.~Recati and S.~Stringari,
Supersolidity in ultracold dipolar gases,
\href{https://doi.org/10.1038/s42254-023-00648-2}
{Nat. Rev. Phys. \textbf{5}, 735 (2023)}.

\bibitem{Sinha_review}
S.~Sinha and S.~Sinha,
Supersolid phases of bosons,
\href{https://doi.org/10.1088/1361-648X/adf6fb}
{J. Phys.: Condens. Matter \textbf{37}, 333001 (2025)}.

\bibitem{Chomaz_review}
L.~Chomaz,
Quantum-stabilized states in magnetic dipolar quantum gases,
\href{https://doi.org/10.48550/arXiv.2504.06221}
{arXiv:2504.06221 [cond-mat.quant-gas]}.

\bibitem{Penrose1956}
O.~Penrose and L.~Onsager,
Bose-Einstein Condensation and Liquid Helium,
\href{https://doi.org/10.1103/PhysRev.104.576}
{Phys. Rev. \textbf{104}, 576 (1956)}.

\bibitem{Gross1957}
E.~P. Gross,
Unified Theory of Interacting Bosons,
\href{https://doi.org/10.1103/PhysRev.106.161}
{Phys Rev. \textbf{106}, 161 (1957)}.

\bibitem{Gross1958}
E.~P. Gross,
Classical theory of boson wave fields,
\href{https://doi.org/10.1016/0003-4916(58)90037-X}
{Ann. Phys. (N.Y.) \textbf{4}, 57 (1958)}.

\bibitem{Thouless1969}
D.~J. Thouless,
The flow of a dense superfluid,
\href{https://doi.org/10.1016/0003-4916(69)90286-3}
{Ann. Phys. \textbf{52}, 403 (1969)}.

\bibitem{Andreev1969}
A.~F. Andreev and I.~M. Lifshitz,
Quantum Theory of Defects in Crystals,
Zh. Eksp. Teor. Fiz. \textbf{56}, 2057 (1969)
[Sov. Phys. JETP \textbf{29}, 1107 (1969)].

\bibitem{Chester1970}
G.~V. Chester,
Speculations on Bose-Einstein Condensation and Quantum Crystals,
\href{https://doi.org/10.1103/PhysRevA.2.256}
{Phys. Rev. A \textbf{2}, 256 (1970)}.

\bibitem{Leggett1970}
A.~J. Leggett,
Can a Solid Be ``Superfluid''?,
\href{https://doi.org/10.1103/PhysRevLett.25.1543}
{Phys. Rev. Lett. \textbf{25}, 1543 (1970)}.

\bibitem{Kirzhnits1971}
D.~A. Kirzhnits and Yu.~A. Nepomnyashchii,
Coherent Crystallization of Quantum Liquid,
Zh. Eksp. Teor. Fiz. \textbf{59}, 2203 (1971)
[Sov. Phys. JETP \textbf{32}, 1191 (1971)].

\bibitem{Pitaevskii1984}
L.~P. Pitaevskii,
Layered structure of superfluid ${}^4\mathrm{He}$ with supercritical motion,
Pis'ma Zh. Eksp. Teor. Fiz. \textbf{39}, 423 (1984)
[JETP Lett. \textbf{39}, 511 (1984)].

\bibitem{Saslow1975}
W.~M. Saslow,
Superfluidity of Periodic Solids,
\href{https://doi.org/10.1103/PhysRevLett.36.1151}
{Phys. Rev. Lett. \textbf{36}, 1151 (1975)}.

\bibitem{Pomeau1994}
Y.~Pomeau and S.~Rica,
Dynamics of a model of supersolid,
\href{https://doi.org/10.1103/PhysRevLett.72.2426}
{Phys. Rev. Lett. \textbf{72}, 2426 (1994)}.

\bibitem{Leggett1998}
A.~J. Leggett,
On the Superfluid Fraction of an Arbitrary Many-Body System at $T = 0$,
\href{https://doi.org/10.1023/B:JOSS.0000033170.38619.6c}
{J. Stat. Phys. \textbf{93}, 927 (1998)}.

\bibitem{Leonard2017}
J.~L\'{e}onard, A.~Morales, P.~Zupancic, T.~Esslinger, and T.~Donner,
Supersolid formation in a quantum gas breaking continuous
translational symmetry,
\href{https://doi.org/10.1038/nature21067}
{Nature (London) \textbf{543}, 87 (2017)}.

\bibitem{Li2017}
J.~Li, J.~Lee, W.~Huang, S.~Burchesky, B.~Shteynas,
F.~\c{C}. Top, A.~O. Jamison, and W.~Ketterle,
A stripe phase with supersolid properties in spin-orbit-coupled
Bose-Einstein condensates,
\href{https://doi.org/10.1038/nature21431}
{Nature (London) \textbf{543}, 91 (2017)}.

\bibitem{Putra2020}
A.~Putra, F.~Salces-C\'{a}rcoba, Y.~Yue, S.~Sugawa, and I.~B. Spielman,
Spatial Coherence of Spin-Orbit-Coupled Bose Gases,
\href{https://doi.org/10.1103/PhysRevLett.124.053605}
{Phys. Rev. Lett. \textbf{124}, 053605 (2020)}.

\bibitem{Chisholm2024}
C.~S. Chisholm, S.~Hirthe, V.~B. Makhalov, R.~Ramos,
R.~Vatr\'{e}, J.~Cabedo, A.~Celi, and L.~Tarruell,
Probing supersolidity through excitations
in a spin-orbit-coupled Bose-Einstein condensate,
\href{https://doi.org/10.48550/arXiv.2412.13861}
{arXiv:2412.13861 [cond-mat.quant-gas]}.

\bibitem{Tanzi2019}
L.~Tanzi, E.~Lucioni, F.~Fam\`{a}, J.~Catani, A.~Fioretti, C.~Gabbanini,
R.~N. Bisset, L.~Santos, and G.~Modugno,
Observation of a Dipolar Quantum Gas with Metastable Supersolid Properties,
\href{https://doi.org/10.1103/PhysRevLett.122.130405}
{Phys. Rev. Lett. \textbf{122}, 130405 (2019)}.

\bibitem{Boettcher2019}
F.~B\"{o}ttcher, J.-N. Schmidt, M.~Wenzel, J.~Hertkorn, M.~Guo, T.~Langen,
and T.~Pfau,
Transient Supersolid Properties in an Array of Dipolar Quantum Droplets,
\href{https://doi.org/10.1103/PhysRevX.9.011051}
{Phys. Rev. X \textbf{9}, 011051 (2019)}.

\bibitem{Chomaz2019}
L.~Chomaz, D.~Petter, P.~Ilzh\"{o}fer, G.~Natale, A.~Trautmann, C.~Politi,
G.~Durastante, R.~M.~W. van Bijnen, A.~Patscheider, M.~Sohmen, M.~J. Mark,
and F.~Ferlaino,
Long-Lived and Transient Supersolid Behaviors in Dipolar Quantum Gases,
\href{https://doi.org/10.1103/PhysRevX.9.021012}
{Phys. Rev. X \textbf{9}, 021012 (2019)}.

\bibitem{Norcia2021}
M.~A. Norcia, C.~Politi, L.~Klaus, E.~Poli, M.~Sohmen,
M.~J. Mark, R.~N. Bisset, L.~Santos, and F.~Ferlaino,
Two-dimensional supersolidity in a dipolar quantum gas,
\href{https://doi.org/10.1038/s41586-021-03725-7}
{Nature (London) \textbf{596}, 357 (2021)}.

\bibitem{Bland2021}
T.~Bland, E.~Poli, C.~Politi, L.~Klaus, M.~A. Norcia, F.~Ferlaino,
L.~Santos, R.~N. Bisset,
Two-Dimensional Supersolid Formation in Dipolar Condensates,
\href{https://doi.org/10.1103/PhysRevLett.128.195302}
{Phys. Rev. Lett. \textbf{128}, 195302 (2022)}.

\bibitem{Trypogeorgos2025}
D.~Trypogeorgos, A.~Gianfrate, M.~Landini, D.~Nigro, D.~Gerace,
I.~Carusotto, F.~Riminucci, K.~W. Baldwin, L.~N. Pfeiffer,
G.~I. Martone, M.~De Giorgi, D.~Ballarini, and D.~Sanvitto,
Emerging supersolidity in photonic-crystal polariton condensates,
\href{https://doi.org/10.1038/s41586-025-08616-9}
{Nature (London) \textbf{639}, 337 (2025)}.

\bibitem{Muszynski2024}
M.~Muszy\'{n}ski \textit{et al.},
Observation of a stripe phase in a spin-orbit coupled exciton-polariton
Bose-Einstein condensate,
\href{https://doi.org/10.48550/arXiv.2407.02406}
{arXiv:2407.02406 [cond-mat.mes-hall]}.

\bibitem{Liebster2025}
N.~Liebster, M.~Sparn, E.~Kath, J.~Duchene, H.~Strobel,
and M.~K. Oberthaler,
Supersolid-like sound modes in a driven quantum gas,
\href{https://doi.org/10.1038/s41567-025-02927-4}
{Nat. Phys. \textbf{21}, 1064 (2025)}.

\bibitem{Sepulveda2010}
N. Sep\'{u}lveda, C. Josserand, and S. Rica,
Superfluid density in a two-dimensional model of supersolid,
\href{https://doi.org/10.1140/epjb/e2010-10176-y}
{Eur. Phys. J. B \textbf{78}, 439 (2010)}.

\bibitem{Roccuzzo2019}
S.~M. Roccuzzo and F.~Ancilotto,
Supersolid behavior of a dipolar Bose-Einstein condensate confined in a tube,
\href{https://doi.org/10.1103/PhysRevA.99.041601}
{Phys. Rev. A \textbf{99}, 041601(R) (2019)}.

\bibitem{Martone2021a}
G.~I. Martone, A.~Recati, and N.~Pavloff,
Supersolidity of cnoidal waves in an ultracold Bose gas,
\href{https://doi.org/10.1103/PhysRevResearch.3.013143}
{Phys. Rev. Res. \textbf{3}, 013143 (2021)}.

\bibitem{Ancillotto2021}
F.~Ancilotto, M.~Barranco, M.~Pi, and L.~Reatto,
Vortex properties in the extended supersolid phase of dipolar Bose-Einstein condensates,
\href{https://doi.org/10.1103/PhysRevA.103.033314}
{Phys. Rev. A \textbf{103}, 033314 (2021)}.

\bibitem{Ripley2023}
B.~T.~E. Ripley, D.~Baillie, and P.~B. Blakie,
Two-dimensional supersolidity in a planar dipolar Bose gas,
\href{https://doi.org/10.1103/PhysRevA.108.053321}
{Phys. Rev. A \textbf{108}, 053321 (2023)}.

\bibitem{Blakie2024}
P.~B. Blakie,
Superfluid fraction tensor of a two-dimensional supersolid,
\href{https://doi.org/10.1088/1361-6455/ad41c1}
{J. Phys. B: At. Mol. Opt. Phys. \textbf{57}, 115301 (2024)}.

\bibitem{Orso2024}
G.~Orso and S.~Stringari,
Superfluid fraction and Leggett Bound in a density-modulated
strongly interacting Fermi gas at zero temperature,
\href{https://doi.org/10.1103/PhysRevA.109.023301}
{Phys. Rev. A \textbf{109}, 023301 (2024)}.

\bibitem{Roccuzzo2020}
S.~M. Roccuzzo, A.~Gallem\'{i}, A.~Recati, and S.~Stringari,
Rotating a Supersolid Dipolar Gas,
\href{https://doi.org/10.1103/PhysRevLett.124.045702}
{Phys. Rev. Lett. 124, 045702 (2020)}.

\bibitem{Gallemi2022}
A.~Gallem\'{i} and L.~Santos,
Superfluid properties of a honeycomb dipolar supersolid,
\href{https://doi.org/10.1103/PhysRevA.106.063301}
{Phys. Rev. A 106, 063301 (2022)}.

\bibitem{Josserand2007}
C.~Josserand, Y.~Pomeau, and S.~Rica,
Coexistence of ordinary elasticity and superfluidity
in a model of a defect-free supersolid,
\href{https://doi.org/10.1103/PhysRevLett.98.195301}
{Phys. Rev. Lett. \textbf{98}, 195301 (2007)}.

\bibitem{Roccuzzo2022}
S.~M. Roccuzzo, A.~Recati, and S.~Stringari,
Moment of inertia and dynamical rotational response of a supersolid dipolar gas,
\href{https://doi.org/10.1103/PhysRevA.105.023316}
{Phys. Rev. A \textbf{105}, 023316 (2022)}.

\bibitem{Sindik2024}
M.~\v{S}indik, T.~Zawi\'{s}lak, A.~Recati, and S.~Stringari,
Sound, Superfluidity, and Layer Compressibility in a Ring Dipolar Supersolid,
\href{https://doi.org/10.1103/PhysRevLett.132.146001}
{Phys. Rev. Lett. \textbf{132}, 146001 (2024)}.

\bibitem{Preti2025}
N.~Preti, N.~Antolini, C.~Drevon, P.~Lombardi, A.~Fioretti,
C.~Gabbanini, G.~Ferioli, G.~Modugno, and G.~Biagioni,
Single-fluid model for rotating annular supersolids
and its experimental implications,
\href{https://doi.org/10.48550/arXiv.2510.26753}
{arXiv:2510.26753 [cond-mat.quant-gas]}.

\bibitem{Tanzi2021}
L.~Tanzi, J.~G. Maloberti, G.~Biagioni, A.~Fioretti,
C.~Gabbanini, and G.~Modugno,
Evidence of superfluidity in a dipolar supersolid
from nonclassical rotational inertia,
\href{https://doi.org/10.1126/science.aba4309}
{Science \textbf{371}, 1162 (2021)}.

\bibitem{Norcia2022}
M.~A. Norcia, E.~Poli, C.~Politi, L.~Klaus, T.~Bland, M.~J. Mark,
L.~Santos, R.~N. Bisset, and F.~Ferlaino,
Can Angular Oscillations Probe Superfluidity in Dipolar Supersolids?,
\href{https://doi.org/10.1103/PhysRevLett.129.040403}
{Phys. Rev. Lett. \textbf{129}, 040403 (2022)}.

\bibitem{Biagioni2024}
G.~Biagioni, N.~Antolini, B.~Donelli, L.~Pezzè, A.~Smerzi, M.~Fattori,
A.~Fioretti, C.~Gabbanini, M.~Inguscio, L.~Tanzi, and G.~Modugno,
Measurement of the superfluid fraction of a supersolid by Josephson effect,
\href{https://doi.org/10.1038/s41586-024-07361-9}
{Nature (London) \textbf{629}, 773 (2024)}.

\bibitem{Kunimi2012}
M.~Kunimi and Y.~Kato,
Mean-field and stability analyses of two-dimensional flowing
soft-core bosons modeling a supersolid,
\href{https://doi.org/10.1103/PhysRevB.86.060510}
{Phys. Rev. B \textbf{86}, 060510(R) (2012)}.

\bibitem{Nilsson2021}
M.~Nilsson Tengstrand, D.~Boholm, R.~Sachdeva, J.~Bengtsson, and S.~M. Reimann,
Persistent currents in toroidal dipolar supersolids,
\href{https://doi.org/10.1103/PhysRevA.103.013313}
{Phys. Rev. A \textbf{103}, 013313 (2021)}.

\bibitem{Mukherjee2025}
K.~Mukherjee, T.~Arnone Cardinale, and S.~M. Reimann,
Selective rotation and attractive persistent currents in antidipolar
ring supersolids,
\href{https://doi.org/10.1103/PhysRevA.111.033304}
{Phys. Rev. A \textbf{111}, 033304 (2025)}.

\bibitem{Gallemi2020}
A.~Gallem\'{i}, S.~M. Roccuzzo, S.~Stringari, and A.~Recati,
Quantized vortices in dipolar supersolid Bose-Einstein-condensed gases,
\href{https://doi.org/10.1103/PhysRevA.102.023322}
{Phys. Rev. A \textbf{102}, 023322 (2020)}.

\bibitem{Sindik2022}
M.~\v{S}indik, A.~Recati, S.~M. Roccuzzo, L.~Santos, and S.~Stringari,
Creation and robustness of quantized vortices in a dipolar supersolid
when crossing the superfluid-to-supersolid transition,
\href{https://doi.org/10.1103/PhysRevA.106.L061303}
{Phys. Rev. A \textbf{106}, L061303 (2022)}.

\bibitem{Schubert2025}
M.~Schubert, K.~Mukherjee, T.~Pfau, and S.~Reimann,
Josephson vortices and persistent current in a double-ring supersolid system,
\href{https://doi.org/10.1103/tl7c-v5bs}
{Phys. Rev. Research \textbf{7}, 033110 (2025)}.

\bibitem{Casotti2024}
E.~Casotti, E.~Poli, L.~Klaus, A.~Litvinov, C.~Ulm, C.~Politi,
M.~J. Mark, T.~Bland, and F.~Ferlaino,
Observation of vortices in a dipolar supersolid,
\href{https://doi.org/10.1038/s41586-024-08149-7}
{Nature (London) \textbf{635}, 327 (2024)}.

\bibitem{Poli2025}
E.~Poli, A.~Litvinov, E.~Casotti, C.~Ulm, L.~Klaus,
M.~J. Mark, G.~Lamporesi, T.~Bland, and F.~Ferlaino,
Synchronization in rotating supersolids,
\href{https://doi.org/10.1038/s41567-025-03065-7}
{Nat. Phys. \textbf{21}, 1820 (2025)}.

\bibitem{Zawislak2025}
T.~Zawi\'{s}lak, M.~\v{S}indik, S.~Stringari, and A.~Recati,
Anomalous Doppler Effect in Superfluid and Supersolid Atomic Gases,
\href{https://doi.org/10.1103/PhysRevLett.134.226001}
{Phys. Rev. Lett. \textbf{134}, 226001 (2025)}.

\bibitem{Martone2018}
G.~I. Martone and G.~V. Shlyapnikov,
Drag force and superfluidity in the supersolid stripe phase
of a spin-orbit-coupled Bose-Einstein condensate,
\href{https://doi.org/10.1134/S004445101811007X}
{Zh. Eksp. Teor. Fiz. \textbf{154}, 985 (2018)}
[\href{https://doi.org/10.1134/S1063776118110146}
{J. Exp. Theor. Phys. \textbf{127}, 865 (2018)}].

\bibitem{Fisher1973}
M.~E. Fisher, M.~N. Barber, and D.~Jasnow,
Helicity Modulus, Superfluidity, and Scaling in Isotropic Systems,
\href{https://doi.org/10.1103/PhysRevA.8.1111}
{Phys. Rev. A \textbf{8}, 1111 (1973)}.

\bibitem{Zhang2016}
Yi-C. Zhang, Z.-Q. Yu, T.~K. Ng, S.~Zhang,
L.~P. Pitaevskii, and S.~Stringari,
Superfluid density of a spin-orbit-coupled Bose gas,
\href{https://doi.org/10.1103/PhysRevA.94.033635}
{Phys. Rev. A \textbf{94}, 033635 (2016)}.

\bibitem{Chen2018}
X.-L. Chen, J.~Wang, Y.~Li, X.-J. Liu and H.~Hu,
Quantum depletion and superfluid density
of a supersolid in Raman spin-orbit-coupled Bose gases,
\href{https://doi.org/10.1103/PhysRevA.98.013614}
{Phys. Rev. A \textbf{98}, 013614 (2018)}.

\bibitem{Martone2021b}
G.~I. Martone, and S.~Stringari,
Supersolid phase of a spin-orbit-coupled
Bose-Einstein condensate: A perturbation approach,
\href{https://doi.org/10.21468/SciPostPhys.11.5.092}
{SciPost Phys. \textbf{11}, 092 (2021)}.

\bibitem{Zhu2012}
Q.~Zhu, C.~Zhang, and B.~Wu,
Exotic superfluidity in spin-orbit coupled Bose-Einstein condensates,
\href{https://doi.org/10.1209/0295-5075/100/50003}
{EPL \textbf{100}, 50003 (2012)}.

\bibitem{Zheng2013}
W.~Zheng, Z.-Q.~Yu, X.~Cui, and H.~Zhai,
Properties of Bose gases with the Raman-induced
spin–orbit coupling,
\href{https://doi.org/10.1088/0953-4075/46/13/134007}
{J. Phys. B \textbf{46}, 134007 (2013)}.

\bibitem{Ozawa2013}
T.~Ozawa, L.~P. Pitaevskii, and S.~Stringari,
Supercurrent and dynamical instability
of spin-orbit-coupled ultracold Bose gases,
\href{https://doi.org/10.1103/PhysRevA.87.063610}
{Phys. Rev. A \textbf{87}, 063610 (2013)}.

\bibitem{Landau1941}
L.~D. Landau,
The theory of superfluidity of helium II,
J. Phys. USSR \textbf{5}, 71 (1941).

\bibitem{Lifshitz_Pitaevskii_book}
E.~M. Lifshitz and L.~P. Pitaevskii,
\textit{Statistical Physics, Part 2} (Pergamon, Oxford, 1980).

\bibitem{Lyu2024}
H.~Lyu, Y.~Chen, Q.~Zhu, and Y.~Zhang,
Supercurrent-carrying supersolid
in spin-orbit-coupled Bose-Einstein condensates,
\href{https://doi.org/10.1103/PhysRevResearch.6.023048}
{Phys. Rev. Research \textbf{6}, 023048 (2024)}.

\bibitem{Geier2021}
K.~T. Geier, G.~I. Martone, P.~Hauke, and S.~Stringari,
Exciting the Goldstone Modes of a Supersolid Spin-Orbit-Coupled Bose Gas,
\href{https://doi.org/10.1103/PhysRevLett.127.115301}
{Phys. Rev. Lett. \textbf{127}, 115301 (2021)}.

\bibitem{Geier2023}
K.~T. Geier, G.~I. Martone, P.~Hauke, W.~Ketterle, and S.~Stringari,
Dynamics of Stripe Patterns in Supersolid Spin-Orbit-Coupled Bose Gases,
\href{https://doi.org/10.1103/PhysRevLett.130.156001}
{Phys. Rev. Lett. \textbf{130}, 156001 (2023)}.

\bibitem{Martone2012}
G.~I. Martone, Y.~Li, L.~P. Pitaevskii, and S.~Stringari,
Anisotropic dynamics of a spin-orbit-coupled
Bose-Einstein condensate,
\href{https://doi.org/10.1103/PhysRevA.86.063621}
{Phys. Rev. A \textbf{86}, 063621 (2012)}.

\bibitem{Lin2011}
Y.-J. Lin, K.~Jimenez-Garcia, and I.~B. Spielman,
Spin-orbit-coupled Bose-Einstein condensates,
\href{https://doi.org/10.1038/nature09887}
{Nature (London) \textbf{471}, 83 (2011)}.

\bibitem{Ho2011}
T.-L. Ho and S.~Zhang,
Bose-Einstein Condensates with Spin-Orbit Interaction,
\href{https://doi.org/10.1103/PhysRevLett.107.150403}
{Phys. Rev. Lett. \textbf{107}, 150403 (2011)}.

\bibitem{Li2012a}
Y.~Li, L.~P. Pitaevskii, and S.~Stringari,
Quantum Tricriticality and Phase Transitions
in Spin-Orbit Coupled Bose-Einstein Condensates,
\href{https://doi.org/10.1103/PhysRevLett.108.225301}
{Phys. Rev. Lett. \textbf{108}, 225301 (2012)}.

\bibitem{Zhou2013_review}
X.~Zhou, Y.~Li, Z.~Cai, and C.~Wu,
Unconventional states of bosons with the synthetic spin-orbit coupling,
\href{https://doi.org/10.1088/0953-4075/46/13/134001}
{J. Phys. B \textbf{46}, 134001 (2013)}.

\bibitem{Zhai2015_review}
H.~Zhai,
Degenerate quantum gases with spin-orbit coupling: a review,
\href{https://doi.org/10.1088/0034-4885/78/2/026001}
{Rep. Prog. Phys. \textbf{78}, 026001 (2015)}.

\bibitem{Li2015_review}
Y.~Li, G.~I. Martone, and S.~Stringari, Spin-Orbit-Coupled Bose-Einstein Condensates,
in \textit{Annual Review of Cold Atoms and Molecules}, Vol.~3,
edited by K. W. Madison, K. Bongs, L. D. Carr, A. M. Rey, H. Zhai
(World Scientific, Singapore, 2015), Chap.~5, pp.~201--250.

\bibitem{Zhang2016_review}
Y.~Zhang, M.~E. Mossman, T.~Busch, P.~Engels, and C.~Zhang,
Properties of spin-orbit-coupled Bose-Einstein condensates,
\href{https://doi.org/10.1007/s11467-016-0560-y}
{Front. Phys. \textbf{11}, 118103 (2016)}.

\bibitem{Martone2023_review}
G.~I. Martone,
Bose-Einstein condensates with Raman-induced
spin-orbit coupling: An overview,
\href{https://doi.org/10.1209/0295-5075/ace2e8}
{Europhys. Lett. \textbf{143}, 25001 (2023)}.

\bibitem{Li2013}
Y.~Li, G.~I. Martone, L.~P. Pitaevskii, and S.~Stringari,
Superstripes and the Excitation Spectrum
of a Spin-Orbit-Coupled Bose-Einstein Condensate,
\href{https://doi.org/10.1103/PhysRevLett.110.235302}
{Phys. Rev. Lett. \textbf{110}, 235302 (2013)}.

\bibitem{SanchezBaena2020}
J.~S\'{a}nchez-Baena, J.~Boronat, and F.~Mazzanti,
Supersolid stripes enhanced by correlations
in a Raman spin-orbit-coupled system,
\href{https://doi.org/10.1103/PhysRevA.101.043602}
{Phys. Rev. A \textbf{101}, 043602 (2020)}.

\bibitem{Wang2010}
C.~Wang, C.~Gao, C.-M.~Jian, and H.~Zhai,
Spin-Orbit Coupled Spinor Bose-Einstein Condensates,
\href{https://doi.org/10.1103/PhysRevLett.105.160403}
{Phys. Rev. Lett. \textbf{105}, 160403 (2010)}.

\bibitem{Wu2011}
C.-J. Wu, I.~Mondragon-Shem, and X.-F. Zhou,
Unconventional Bose–Einstein Condensations from Spin-Orbit Coupling,
\href{https://doi.org/10.1088/0256-307X/28/9/097102}
{Chin. Phys. Lett. \textbf{28}, 097102 (2011)}.

\bibitem{Xia2023}
W.-L. Xia, L.~Chen, T.-T. Li, Y.~Zhang, and Q.~Zhu,
Metastable supersolid in spin-orbit-coupled Bose-Einstein condensates,
\href{https://doi.org/10.1103/PhysRevA.107.053302}
{Phys. Rev. A \textbf{107}, 053302 (2023)}.

\bibitem{Pitaevskii_Stringari_book}
L.~P. Pitaevskii and S.~Stringari,
\textit{Bose-Einstein Condensation and Superfluidity}
(Oxford University Press, Oxford, 2016).

\bibitem{Pethick_Smith_book}
C.~J. Pethick and H.~Smith,
\textit{Bose-Einstein Condensation in Dilute Gases}
(Cambridge University Press, Cambridge, 2008), 2nd Edition.

\bibitem{Shelykh2010}
I.~A. Shelykh, A.~V. Kavokin, Y.~G. Rubo, T.~C.~H. Liew, and G.~Malpuech,
Polariton polarization-sensitive phenomena in planar semiconductor microcavities,
\href{https://doi.org/10.1088/0268-1242/25/1/013001}
{Semicond. Sci. Technol. \textbf{25}, 013001 (2010)}.

\bibitem{Hamner2015}
C.~Hamner, Y.~Zhang, M.~A. Khamehchi, M.~J. Davis, and P. Engels,
Spin-Orbit-Coupled Bose-Einstein Condensates
in a One-Dimensional Optical Lattice,
\href{https://doi.org/10.1103/PhysRevLett.114.070401}
{Phys. Rev. Lett. \textbf{114}, 070401 (2015)}.

\bibitem{Castin_review}
Y.~Castin,
Bose-Einstein Condensates in Atomic Gases: Simple Theoretical Results,
in \textit{Coherent atomic matter waves. Les Houches
- Ecole d’Ete de Physique Theorique},
edited by R. Kaiser, C. Westbrook, and F. David
(Springer, Berlin, Heidelberg, 2001), Vol. 72, pp. 1-136.

\bibitem{Zheng2012}
W.~Zheng and Z.~Li,
Collective modes of a spin-orbit-coupled Bose-Einstein condensate:
A hydrodynamic approach,
\href{https://doi.org/10.1103/PhysRevA.85.053607}
{Phys. Rev. A \textbf{85}, 053607 (2012)}.

\bibitem{Khamehchi2014}
M.~A. Khamehchi, Y.~Zhang, C.~Hamner, T.~Busch, and P.~Engels,
Measurement of collective excitations in a spin-orbit-coupled
Bose-Einstein condensate,
\href{https://doi.org/10.1103/PhysRevA.90.063624}
{Phys. Rev. A \textbf{90}, 063624 (2014)}.

\bibitem{Ji2015}
S.-C. Ji, L.~Zhang, X.-T. Xu, Z.~Wu, Y.~Deng, S.~Chen, and J.-W. Pan,
Softening of Roton and Phonon Modes in a Bose-Einstein Condensate
with Spin-Orbit Coupling,
\href{https://doi.org/10.1103/PhysRevLett.114.105301}
{Phys. Rev. Lett. \textbf{114}, 105301 (2015)}.

\bibitem{Li2016}
J.~Li, W.~Huang, B.~Shteynas, S.~Burchesky, F.~\c{C}. Top,
E.~Su, J.~Lee, A.~O. Jamison, and W.~Ketterle,
Spin-Orbit Coupling and Spin Textures in Optical Superlattices,
\href{https://doi.org/10.1103/PhysRevLett.117.185301}
{Phys. Rev. Lett. \textbf{117}, 185301 (2016)}.

\bibitem{Martone2014}
G.~I. Martone, Y.~Li, and S.~Stringari,
Approach for making visible and stable stripes in a spin-orbit-coupled
Bose-Einstein superfluid,
\href{https://doi.org/10.1103/PhysRevA.90.041604}
{Phys. Rev. A \textbf{90}, 041604(R) (2014)}.

\bibitem{Martone2015}
G.~I. Martone,
Visibility and stability of superstripes in a spin-orbit-coupled
Bose-Einstein condensate,
\href{https://doi.org/10.1140/epjst/e2015-02386-x}
{Eur. Phys. J. Special Topics \textbf{224}, 553 (2015)}.

\bibitem{deHond2022}
J.~de Hond, J.~Xiang, W.~C. Chung, E.~Cruz-Col\'{o}n, W.~Chen,
W.~C. Burton, C.~J. Kennedy, and W.~Ketterle,
Preparation of the Spin-Mott State: A Spinful Mott Insulator 
of Repulsively Bound Pairs,
\href{https://doi.org/10.1103/PhysRevLett.128.093401}
{Phys. Rev. Lett. \textbf{128}, 093401 (2022)}.

\end{thebibliography}
\end{document}